\documentclass[twocolumn,numberedappendix, twocolappendix, apj]{openjournal}
\usepackage{hyperref}
\usepackage{graphicx}
\usepackage{amsmath,amssymb}
\usepackage{lastpage}
\usepackage[all]{hypcap}
\usepackage[T1]{fontenc}
\usepackage{float}
\usepackage{subfigure}
\usepackage{afterpage}
\usepackage[usenames]{color}
\usepackage{mathrsfs}
\usepackage{upgreek}
\usepackage{color}
\usepackage[dvipsnames]{xcolor}
\usepackage{longtable}
\usepackage{threeparttable}
\usepackage{multirow}
\usepackage{placeins}
\usepackage{bm}
\usepackage{fontawesome5}
\usepackage{enumitem}
\usepackage{cool}
\usepackage{float}
\makeatletter
\let\newfloat\newfloat@ltx
\makeatother
\usepackage{algorithm}
\usepackage{algpseudocode}
\usepackage{fontawesome5}
\usepackage{soul}

\newcommand{\github}[1]{\href{#1}{\faGithubSquare}}
\newcommand{\roxygithub}{\github{https://github.com/DeaglanBartlett/roxy}}
\newcommand{\unif}{\texttt{unif}}
\newcommand{\prof}{\texttt{prof}}
\newcommand{\mnr}{\texttt{mnr}}

\newcommand{\dd}{\mathrm{d}}
\newcommand{\sint}{\sigma_\text{int}}
\DeclareMathOperator{\var}{Var}
\DeclareMathOperator{\cov}{Cov}
\DeclareMathOperator{\atantwo}{atan2}
\DeclareMathOperator{\diag}{diag}
\newcommand{\xt}{x_{\rm t}}
\newcommand{\yt}{y_{\rm t}}
\newcommand{\xo}{x_{\rm o}}
\newcommand{\yo}{y_{\rm o}}
\newcommand{\iu}{\mathrm{i}\mkern1mu}
\newcommand{\bxt}{\bm{x}_{\rm t}}

\def\app#1#2{%
  \mathrel{%
    \setbox0=\hbox{$#1\sim$}%
    \setbox2=\hbox{%
      \rlap{\hbox{$#1\propto$}}%
      \lower1.1\ht0\box0%
    }%
    \raise0.25\ht2\box2%
  }%
}

\date{Accepted XXX. Received YYY; in original form ZZZ}

\begin{document}
\label{firstpage}

\title[Marginalised Normal Regression]{Marginalised Normal Regression: unbiased curve fitting in the presence of x-errors}

\author{Deaglan J. Bartlett$^{*}$}
\email{$^*$deaglan.bartlett@iap.fr}
\affiliation{CNRS \& Sorbonne Universit\'e, Institut d’Astrophysique de Paris (IAP), UMR 7095, 98 bis bd Arago, F-75014 Paris, France}
\author{Harry Desmond$^{\dagger}$}
\email{$^{\dagger}$harry.desmond@port.ac.uk}
\affiliation{Institute of Cosmology \& Gravitation, University of Portsmouth, Dennis Sciama Building, Portsmouth, PO1 3FX, UK}

\begin{abstract}
The history of the seemingly simple problem of straight line fitting in the presence of both $x$ and $y$ errors has been fraught with misadventure, with statistically ad hoc and poorly tested methods abounding in the literature.
The problem stems from the emergence of latent variables describing the ``true'' values of the independent variables, the priors on which have a significant impact on the regression result. By analytic calculation of maximum 
\textit{a posteriori}
values and biases, and comprehensive numerical mock tests, we assess the quality of possible priors. In the presence of intrinsic scatter, the only prior that we find to give reliably unbiased results in general is a mixture of one or more Gaussians with means and variances determined as part of the inference.
We find that a single Gaussian is typically sufficient and dub this model
\emph{Marginalised Normal Regression} (\mnr).
We illustrate the necessity for \mnr{} by comparing it to alternative methods on 
an important linear relation in cosmology,
and extend it to nonlinear regression
and an arbitrary covariance matrix linking $x$ and $y$.
We publicly release a Python/Jax implementation of \mnr{}
and its Gaussian mixture model extension that is
coupled to Hamiltonian Monte Carlo for efficient sampling, which we call \textsc{roxy} (\emph{Regression and Optimisation with X and Y errors}) \roxygithub.
\end{abstract}

\section{Introduction}\label{sec:intro}


The problem of determining the parameters of a functional fit to a data set is perhaps the most basic in science. Assuming no prior knowledge on the parameters this is achieved by maximising the likelihood of the data given the fit and the error model, and deriving confidence intervals on the parameters from the isocontours of log-likelihood. With uncorrelated Gaussian uncertainties this is trivial in case only the dependent variable (``$y$'') is uncertain---up to a constant the log-likelihood is the 
negative
sum over points of the squared residuals from the functional expectation divided by the squared uncertainty.

Unfortunately, the independent variable of the regression (``$x$'') is itself often uncertain. Perhaps surprisingly, this greatly increases the complexity of the inference regardless of the functional form being fit. This has led to much confusion in the literature, with researchers often seeming to despair of there being a correct solution and instead running a battery of ad hoc algorithms and taking the spread in their results as a measure of systematic error
\cite{Isobe, Kelly_2011}. 
For example, 
focusing on the case of linear regression,
\citet{ODR} suggest using the orthogonal distance regression (ODR) method, whereby one minimises the (weighted) sum of squares of the orthogonal distance between the observed data points and the line.
A popular method for dealing with the asymmetry between the dependent and independent variable has been the so-called ``forward'' and ``backward'' methods, whereby one performs linear fits of $y$ vs $x$ and then $x$ vs $y$, 
taking account the uncertainty on the dependent variable only in each case. The results are then
combined in somewhat ad hoc ways \citep[see, e.g.][]{Isobe_1990,Feigelson_1992}, including the ``bisector'' method where one chooses the parameters of the straight line that bisects these two fits.
Another method, dubbed BCES \citep{Akritas_1996} is a weighted least squares estimator that extends the ordinary least-squares (OLS) method. This has been criticised 
by \citet{Tremaine_2002}
due to its being independent of the errors on the dependent variable and suffering from the potentially large impact of a few low-error measurements, poor performance on mock data and a potentially numerically unstable denominator.

An important consideration is the way in which the regression method treats intrinsic scatter in the fitted relation. This describes variability in $y$ at fixed $x$ (or $x$ at fixed $y$) beyond that accountable for by the uncertainties in the measurements, and may be due to correlation of $x$ and/or $y$ with one or more additional variables. Intrinsic scatter commonly arises when projecting the parameter spaces of physical systems into two dimensions, and cannot be reduced by performing more measurements.
To allow for intrinsic scatter,
\citet{Tremaine_2002} advocate the minimisation of a $\chi^2$ function, where the sum of square residuals of the function are weighted by the sum in quadrature of the $y$ errors,  a variable characterising the intrinsic scatter, and the product of the gradient of the line and the $x$ errors 
\citep[sometimes called the ``Nukers'' method; see also][]{Novak_2006}.
This adjusted $\chi^2$ fitting procedure was incorporated as an extension to the \textsc{mpift} package \citep{mpfit} by \citet{Williams_2010}\footnote{\url{http://purl.org/mike/mpfitexy}}, where the intrinsic scatter is adjusted to give a reduced $\chi^2$ of unity.

In cases where the methods described above have been tested against mock data, they have been found to be biased \citep[e.g.][]{Aguado-Barahona_2019,Aguado-Barahona_2022}. 
In addition, some of them assume homoscedasticity (constant errors among the data points), no intrinsic scatter (i.e. perfect correlation between the underlying $x$ and $y$ values) or a linear underlying relation, or provide point estimates of the regression parameters without confidence intervals.
As these methods are statistically unjustified we do not investigate them here.
For in-depth treatises on measurement error models and their impact on regression see~\citet{Fuller_book, Carroll_book, Buonaccorsi_book}. 

From a Bayesian perspective the setup of the problem is quite simple (see Section~\ref{sec:setup}), and sees the emergence of
new degrees of freedom in the hierarchical model describing the ``true'' $x$ values $\bm{\xt}$, from which the ``observed'' values, $\bm{\xo}$, are derived by scattering according to the error model.
The distribution of $\bm{\xt}$ values underlies the inference but can only be inferred through its effect on the observed $\bm{\xo}$ and $\bm{\yo}$ values. The likelihood of the data given the $\bm{\xt}$ and fit parameters is simple to write down (Eq.~\ref{eq:likelihood}). When the intrinsic scatter $\sint$ in the relation is assumed to be 0 (i.e. the true model $y$ value, $\bm{\yt}$, is deterministically related to $\bm{\xt}$), the maximum of this likelihood is at the true parameter values. This means that the profile likelihood (equivalent to marginalising over $\bm{\xt}$ given delta-function priors at their maximum-likelihood values) is unbiased. The neglect of the uncertainties of $\bm{\xt}$ can however lead the profile likelihood to underestimate the uncertainties on the fit parameters. This can be overcome by marginalising over $\bm{\xt}$ with a Jeffreys prior \citep{Jeffreys_1946}, which is reparametrisation invariant and hence removes volume effects. 
By contrast, marginalising over $\bm{\xt}$ with a uniform prior does introduce volume effects such that the maximum of the marginal likelihood is not at the maximum of Eq.~\ref{eq:likelihood}, biasing the inference. This parallels investigations into methods of marginalising over other types of nuisance parameter, as explored for example in \citet{Berger, Hadzhiyska_2023}.

The problem arises when the intrinsic scatter $\sint$ between $\bm{\xt}$ and $\bm{\yt}$ is assumed not to be 0, putting a free parameter in the \emph{denominator} of the exponential likelihood. In this case, we will show that the maximum of the likelihood is \emph{not} at the true parameter values, so that both the profile likelihood and the one that is
obtained by marginalisation over $\bm{\xt}$ with a Jeffreys prior are biased.
This is not a small effect: the maximum-likelihood values of $\sint^2$ are \emph{negative} until the true $\sint^2$ exceeds several times the other variance scales in the problem. Marginalising over $\bm{\xt}$ with a uniform prior produces a marginal likelihood with a much less biased $\sint$ value at its maximum, but biases in both $\sint$ and the parameters describing the shape of the function being fit remain.
A significant part of the paper is devoted to investigating and quantifying these effects.

How then to achieve an unbiased inference in the presence of intrinsic scatter? One way to think about the problem is to realise that if the parameters of the functional fit are to be inferred correctly, so too must be the $\bm{\xt}$ values. The distribution of these is however not known, and typically none of the marginalisation methods considered above have sufficient flexibility to capture that distribution \textit{a posteriori}.
The problem then is how to create a prior $\bm{\xt}$ distribution that is sufficiently flexible to capture the true distribution but at the same time is computationally tractable given the limited data available. This is studied in~\citet{Gull_1989,Muller_Erkanli_West,Carroll_1999,Schafer_2001,Huang_2006,Kelly_2007,Carroll_2009,Sarkar_2014}, where finally emerge effective solutions.
There are two classes of approach, so-called parametric and nonparametric (or semi-parametric) models. In the former one proposes a parametrised form for the prior on the $\bm{\xt}$ distribution, and those parameters are marginalised over in constraining the parameters of interest. The latter attempts to overcome the rigidity of fixed functional forms by adopting a fully general $\bm{\xt}$ distribution, which may be achieved by histogramming or kernel density estimation, or modern machine learning methods such as neural networks or Gaussian processes. The parametric approach offers stability and interpretability at the cost of potential model misspecification, while the nonparametric approach offers robustness to bias at the cost of potential intractability of the inference if the size of the dataset is insufficient.

A particularly convenient class of parametric model is a Gaussian mixture model (GMM)
which we will show to offer a general-purpose solution to the inference problem across the range of potential scientific applications.
Indeed, we will show that in almost all cases a \emph{single} Gaussian is sufficient to yield unbiased inference, even in cases where the true $\bm{\xt}$ distribution is far from Gaussian.
We show analytically that the bias of this method is exactly zero in the limit of infinite sample size and constant $x$ and $y$ errors, irrespective of the underlying $\bm{\xt}$ distribution, and through an exhaustive set of numerical experiments that the bias very rarely exceeds 2$\sigma$. We dub this method ``\emph{Marginalised Normal Regression} (\mnr)'', and implement it to fit non-linear functions using non-diagonal covariance matrices. We publicly release an efficient 
Python/Jax \citep{jax2018github}
implementation of our method powered by Hamiltonian Monte Carlo, which we call
\textsc{roxy} (\emph{Regression and Optimisation with X and Y errors}).\footnote{\url{https://github.com/DeaglanBartlett/roxy}} This also offers the capability of fitting more than one Gaussian in cases where that alleviates the residual bias, as we discuss further below.

Our 
GMM
algorithm is similar to the method of~\citet{Kelly_2007}, which has an IDL\footnote{\url{https://github.com/wlandsman/IDLAstro/blob/master/pro/linmix_err.pro}} and Python\footnote{\url{https://github.com/jmeyers314/linmix}} implementation. 
This method has also been implemented in the R language with an extension to multiple variables (Linear Regression by Gibbs Sampling - \textsc{lrgs};~\citealt{Mantz_2016}) and with time-dependent parameters (LInear Regression in Astronomy - \textsc{lira};~\citealt{Sereno_2016}).
The primary advantages of \textsc{roxy} over these are efficiency (leveraging automatic differentiation and Hamiltonian Monte Carlo), flexibility (offering multiple sets of hyperpriors, the ability to automatically determine the optimum number of Gaussians, and the ability to fit nonlinear functions) and ease of use, as well as small algorithmic upgrades such as an ordering of the means of multiple Gaussians to remove the degeneracies between them.
Further comparison with the method of~\citeauthor{Kelly_2007} may be found in Sec.~\ref{sec:applications}.

This paper is structured as follows. To begin, we focus for simplicity on linear regression. In Section~\ref{sec:setup} we describe the setup of the problem and derive the likelihood of the data as a function of the straight line parameters and $\bm{\xt}$ as well as the marginal likelihood when integrating out $\bm{\xt}$ under various priors. We analytically calculate the properties of these marginal likelihoods under necessary simplifying assumptions (most importantly homoscedasticity),
referring to Section~\ref{app} for detailed derivations. In Section~\ref{sec:mock} we generate mock datasets with characteristics spanning the range likely to be encountered in real-world applications and fit them with the various likelihoods to quantify the biases they induce. This will show that \mnr{} is effectively never biased, while the other models almost always are. We also provide a comparison with the case of multiple Gaussians,
demonstrating that one Gaussian is almost always sufficient for unbiased regression and that adding further Gaussians does not generically reduce bias.
In Section~\ref{sec:applications} we apply the models to a
real-world test case in cosmology to illustrate the kinds of differences they produce. 
In Section~\ref{sec:nonlinear} we extend the models to nonlinear functions and an arbitrary covariance matrix, 
and in Section~\ref{sec:roxy} we present our implementation in the public \textsc{roxy} program. Section~\ref{sec:conc} concludes.

\section{The problem and its possible solutions}\label{sec:setup}

In this section we set up the seemingly simple problem of fitting a straight line to data, under the assumption of uncorrelated Gaussian errors on the $x$ and $y$ values. We present three methods for determining the parameters of the function: the slope $A$, the intercept, $B$, and the intrinsic scatter in the $y$ direction, $\sigma_{\rm int}$. We generalise this problem to an arbitrary function and covariance matrix in Section~\ref{sec:nonlinear}. 

\subsection{Setup}

We will consider the problem of assuming that the true values of $\bm{y}_{\rm t} = (y_{\rm t}^{(1)}, \ldots , y_{\rm t}^{(N)})$ are related to the true values of $\bm{x}_{\rm t} = (x_{\rm t}^{(1)}, \ldots , x_{\rm t}^{(N)})$ as 
\begin{equation}
	\label{eq:straight line}
	y_{\rm t}^{(i)} = A x_{\rm t}^{(i)} + B,
\end{equation}
We will denote $\bm{\theta}$ as the vector of parameters that we wish to infer. This will include $A$ and $B$, but potentially other parameters too, as outlined in the rest of this section.
Of course, we do not observe the true values, but the set of observed values $\bm{x}_{\rm o}$ and $\bm{y}_{\rm o}$ 
which we assume are Gaussian distributed about the true values, with uncertainties $\sigma_x^{(i)}$ and  $\sigma_y^{(i)}$ in the $x$ and $y$ directions for point $i$, respectively. We can then write the likelihood for some observed $\bm{x}_{\rm o}$ and $\bm{y}_{\rm o}$ given $A$, $B$ and $\bm{x}_{\rm t}$ to be
\begin{equation}
	\label{eq:likelihood}
	\begin{split}
	 	\mathcal{L} & \left(\bm{x}_{\rm o},  \bm{y}_{\rm o} | \bm{x}_{\rm t}, \bm{\theta} \right) =
		\prod_i
	 	\frac{1}{2 \pi \sigma_x^{(i)} \sqrt{\sigma_y^{(i)}{}^2 + \sigma_{\rm int}^2} } \\
	 	& \times \exp \left( - \frac{\left( x_{\rm o}^{(i)} - x_{\rm t}^{(i)} \right)^2}{2 {\sigma_x^{(i)}}^2} \right)
	 	\exp \left( - \frac{\left( y_{\rm o}^{(i)} - A x_{\rm t}^{(i)} - B \right)^2}{2 \left({\sigma_y^{(i)}}{}^2 + \sigma_{\rm int}^2 \right)} \right),
	 \end{split}
\end{equation}
where we have included an additional intrinsic scatter, $\sigma_{\rm int}$, which is not captured by the measurement uncertainties and is another parameter we wish to infer.
In a Bayesian context, we wish to use the likelihood to determine the posterior of $\bm{x}_{\rm t}$ and $\bm{\theta}$ given $\bm{x}_{\rm o}$ and $\bm{y}_{\rm o}$,
\begin{equation}
	P \left( \bm{x}_{\rm t}, \bm{\theta} | \bm{x}_{\rm o},  \bm{y}_{\rm o} \right) = \frac{p(\bm{x}_{\rm t}, \bm{\theta})  \mathcal{L}\left(\bm{x}_{\rm o},  \bm{y}_{\rm o} | \bm{\theta}, \bm{x}_{\rm t}  \right) }{Z },
\end{equation}
where $p(\bm{x}_{\rm t},\bm{\theta})$ is the prior probability of $\bm{x}_{\rm t}$ and $\bm{\theta}$, and $Z$ is the evidence. Given $p(\bm{x}_{\rm t},\bm{\theta})$, one can obtain a numerical approximation to the posterior through Markov chain Monte Carlo (MCMC). Note that, in general, one must sample both $\bm{x}_{\rm t}$ and $\bm{\theta}$  since the likelihood is a function of both,
although in certain cases one can analytically marginalise over $\bm{x}_{\rm t}$ and thus sample only $\bm{\theta}$. 

We now arrive at the crucial question: what prior to impose on $\bxt$? In the following subsections we consider an (improper) infinite uniform, a delta-function at $\bm{\xt}$ values that maximise Eq.~\ref{eq:likelihood} (giving a profile likelihood over the $\bm{\xt}$, which we find to give very similar results to a Jeffreys prior), and a Gaussian hyperprior with mean and width to be inferred.

\subsection{Infinite uniform marginalisation}\label{sec:unif}

A common choice \citep{Agostini} is to give $\bm{\xt}$ (as well as $A$, $B$ and $\sint$) infinite uniform priors, such that $p(\bm{x}_{\rm t}, \bm{\theta})$ is independent of $\bm{x}_{\rm t}$.
The motivation for this choice is that it is seemingly uninformative and maximally agnostic about the $\bm{\xt}$ values, and hence might be thought least likely to introduce a bias if a more restrictive prior on $\bm{\xt}$ is inaccurate. We will see that this is incorrect.

In this case, we can trivially marginalise over $\bm{x}_{\rm t}$ to obtain
\begin{equation}
	\label{eq:posterior uniform straight line}
	\begin{split}
		&P_{\rm U}  (\bm{\theta} | \bm{x}_{\rm o}, \bm{y}_{\rm o}	) 
		\propto
		\int  \mathcal{L}\left(\bm{x}_{\rm o},  \bm{y}_{\rm o} | \bm{x}_{\rm t}, \bm{\theta} \right) \dd^N x_{\rm t} \\
		&= \prod_i \frac{1}{\sqrt{2\pi \left( A^2 {\sigma_x^{(i)}}^2 + {s_y^{(i)}}^2 \right)}} \exp \left( - \frac{\left(A x_{\rm o}^{(i)} + B - y_{\rm o} \right)^2}{2 \left( A^2 {\sigma_x^{(i)}}^2 + {s_y^{(i)}}^2 \right)}\right),
	\end{split}
\end{equation}
where we have introduced ${s_y^{(i)}}^2 \equiv {\sigma_y^{(i)}}^2 + \sigma_{\rm int}^2$ for notational convenience. We denote this likelihood \unif.

Referring to Section~\ref{sec:app uniform} for the full derivation, we find that the 
maximum \textit{a posteriori}
parameter values in the case of universal errors on $x$ and $y$, i.e. $\sigma_x^{(i)} = \sigma_x \, \forall i$ and $\sigma_y^{(i)} = \sigma_y \, \forall i$ are
\begin{align}
    \hat{A} &= \frac{\cov \left(\bm{\xo}, \bm{\yo} \right)}{\var \left( \bm{\xo} \right)}, \\
    \hat{B} &= \frac{\langle \bm{\xo} ^2 \rangle \langle \bm{\yo} \rangle - \langle \bm{\xo} \bm{\yo} \rangle \langle \bm{\xo} \rangle}{\var \left( \bm{\xo} \right)}, \\
    \hat{s}^2 &= 
    \var \left( \bm{\yo} \right)
    - \frac{\cov \left(\bm{\xo}, \bm{\yo} \right)^2}{\var \left( \bm{\xo} \right)^2} \left( \var \left( \bm{\xo} \right) + \sigma_x^2 \right),
\end{align}
where the notation is defined at the start of Section~\ref{app}. In Section~\ref{sec:app uniform} we then compute the difference between these values and the true values in the limit of an infinite sample size and find that these estimates are biased by an amount 
\begin{align}
    \delta \hat{A} &= - \frac{\tilde{A} \sigma_x^2}{\var \left( \bm{\xt} \right) + \sigma_x^2}, \\
    \delta  \hat{B} &= \frac{\tilde{A} \sigma_x^2 \langle \bm{\xt} \rangle}{\var \left( \bm{\xt} \right) + \sigma_x^2}, \\
    \delta \hat{s}^2 &= \frac{\tilde{A}^2 \var \left( \bm{\xt} \right) \sigma_x^4}{\left(\var \left( \bm{\xt} \right) + \sigma_x^2\right)^2},
\end{align}
where a tilde denotes the true values.
Thus, for positive $\tilde{A}$, we would underestimate the gradient and for all non-zero values of $\tilde{A}$ we overestimate the intrinsic scatter. The sign of the bias on $\delta \hat{B}$ is equal to the sign of the product of $\tilde{A}$ and the mean of the underlying true $x$ distribution. 
This clearly demonstrates that the \unif{} method is unreliable. We note that variants of this method have been proposed, e.g. finite uniform priors \citep{Hee_2016}, but it is beyond the scope of our work to investigate them in detail.

\subsection{Profile likelihood}\label{sec:prof}

An alternative is to fix $\bm{\xt}$ to the values that maximise the likelihood Eq.~\ref{eq:likelihood} as a function of $A$, $B$ and $\sint$. The attraction of this approach is that the maximum of this profile likelihood is by construction at the same point in parameter space as the maximum of the full likelihood Eq.~\ref{eq:likelihood}, which might be hoped to be unbiased (we will show that in general it is not). This is in contrast to any of the other methods we discuss, where marginalising out $\bm{\xt}$ introduces a volume effect that shifts the maximum-posterior point away from the maximum-likelihood point.

Maximising Eq.~\ref{eq:likelihood} with respect to each of the $\xt^{(i)}$ values gives maximum-likelihood $\bm{\xt}$ positions of 
\begin{equation}
    \hat{x}_{\rm t}^{(i)} =
     \frac{{s_y^{(i)}}^2}{{s_y^{(i)}}^2 + A^2 {\sigma_x^{(i)}}^2} \xo^{(i)} 
    + \frac{A {\sigma_x^{(i)}}^2}{{s_y^{(i)}}^2 + A^2 {\sigma_x^{(i)}}^2} \left( \yo^{(i)} - B \right).
\end{equation}
One can substitute these values back into Eq.~\ref{eq:likelihood} to obtain the profile likelihood for $A$, $B$ and $\sigma_{\rm int}$,
\begin{equation}
    \label{eq:profile likelihood}
    \begin{split}
	 \mathcal{L}  &\left(\bm{x}_{\rm o},  \bm{y}_{\rm o} | \hat{\bm{x}}_{\rm t}, \bm{\theta} \right) = \\
        &\prod_i
        \frac{1}{\sqrt{2 \pi {s_y^{(i)}}^2}}
        \exp \left( - \frac{\left(\yo^{(i)} - A  \xo^{(i)} - B \right)^2}{2 \left(A ^2 {\sigma_x^{(i)}}^2 + {s_y^{(i)}}^2 \right)} \right),
    \end{split}
\end{equation}
which can then be maximised or sampled to infer the parameters of the fit. We denote this likelihood \prof.

We study this likelihood analytically in Section~\ref{sec:app profile}. Simplifying again to the case of constant $\sigma_x$ and $\sigma_y$, we find that there does not exist a single closed-form expression for all the parameters at maximum-likelihood. Instead, one must solve a cubic equation for $\hat{s}^2$ (see Section~\ref{sec:cubic properties} and Algorithm~\ref{alg:cubic real roots} for a discussion of the real analytic roots of this equation) and the desired root depends on the data used.
Moreover, in some cases the maximum likelihood solution would be at $\hat{s}^2 < 0$, which is clearly unphysical. Thus, not only does one need to consider the turning points of the likelihood, but also the case with $\sigma_{\rm int} = 0$ separately. We will find that in many cases this is the maximum likelihood solution for $\hat{s}^2$, and hence the profile likelihood often severely underestimates the intrinsic scatter.
The differences between the maximum-likelihood and true values of the parameters also do not have a closed-form analytic expression; these are however 
non-zero in all the parameters,
showing that the profile likelihood is biased.

In practice the Jeffreys prior $\sqrt{|\mathcal{I}|}$ \citep{Jeffreys_1946}, where $\mathcal{I}$ is the Fisher information matrix, produces very similar results to the profile likelihood (see also \citealt{Hadzhiyska_2023}). The Jeffreys prior is an attractive choice in the absence of informative prior information because it is reparametrisation invariant, and consequently eliminates volume effects deriving from marginalising over nuisance parameters such as $\bm{\xt}$. The profile likelihood is by construction devoid of volume effects, and gives only slightly smaller posterior widths for the other parameters than Jeffreys marginalisation.
Just like the \prof{} method,
the Jeffreys prior does not produce an unbiased result because in the presence of intrinsic scatter the maximum of the likelihood Eq.~\ref{eq:likelihood} does not occur at the true parameter values. 

\subsection{Gaussian hyperprior}\label{sec:MNR}

\begin{figure}
    \centering
    \includegraphics[width=\columnwidth, trim={0.5cm 1.8cm 0.5cm 1.8cm},clip]{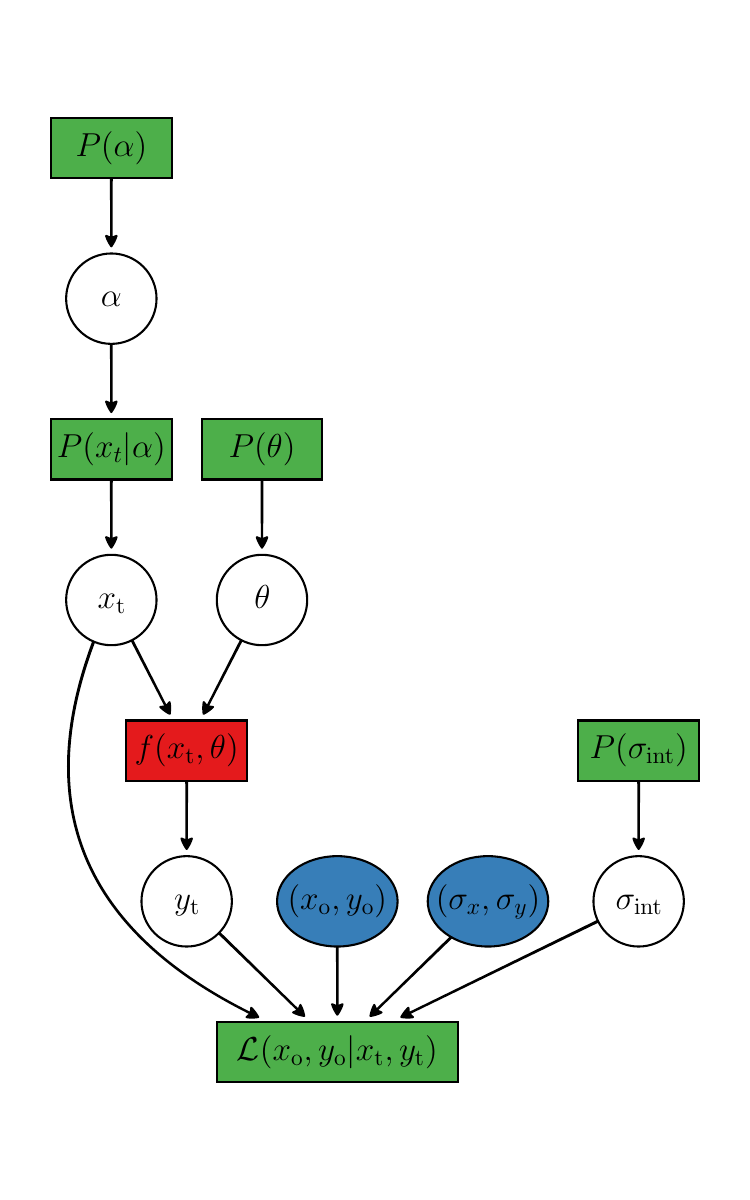}
    \caption{Directed acyclic graph representation of the Bayesian hierarchical model for a functional fit $f$ in the presence of uncertainties in both the independent and dependent variable. Green boxes represent probability distributions, red boxes deterministic relations, blue ellipses observable quantities and white circles variables. For \mnr, $\alpha$ contains the mean and width of the Gaussian hyperprior; for a linear fit, $\theta$ contains the slope and intercept. 
    }
    \label{fig:BHM}
\end{figure}
 
The above likelihoods have in common that they extract no knowledge about the $\bm{\xt}$ values from the data set itself, but rather marginalise over them before the data has been considered. (The Fisher matrix appearing in the Jeffreys prior depends on the \textit{possible} data values, but not on the actual measurements.) This is undesirable: would one think, for example, that the $\bm{\xt}$ values of a dataset where $0<x<5$ (and small $\sigma_x$) could lie anywhere in [$-\infty,\infty$]? The alternative, which is key to removing the biases explored above, is to learn about the $\bm{\xt}$ distribution from the data set itself. In the Bayesian hierarchical model given in Fig.~\ref{fig:BHM}, this involves inferring both the parameters of the equation relating $\bm{\xt}$ and $\bm{\yt}$, $\bm{\theta}$, but also the parameters of the prior on $\bm{\xt}$, $\bm{\alpha}$.

The simplest nontrivial implementation of this is to give $\bm{\xt}$ a Gaussian distribution whose mean $\mu$ and width $w$ are free parameters to be inferred from the data. This yields
\begin{equation}
    \begin{split}
        P & \left( \bm{\theta}, \bm{\xt} \vert \bm{\xo}, \bm{\yo} \right)
        = \prod_i \frac{1}{\sqrt{2\pi w^2}} \exp \left( - \frac{\left(\xt^{(i)} - \mu \right)^2}{2 w^2} \right) \\
        &\times \frac{1}{\sqrt{2\pi {\sigma_x^{(i)}}^2}} \exp \left( - \frac{\left( \xt^{(i)} - \xo^{(i)} \right)^2}{2 {\sigma_x^{(i)}}^2} \right) \\
        &\times \frac{1}{\sqrt{2\pi {s_y^{(i)}}^2}} \exp \left( - \frac{\left( A \xt^{(i)} + B - \yo^{(i)} \right)^2}{2 {s_y^{(i)}}^2} \right).
    \end{split}
\end{equation}
Although one could jointly sample the $\bm{\xt}$ values alongside $\bm{\theta}$, in the majority of the cases we do not wish to find the $\bm{\xt}$ values themselves, so we can analytically marginalise over them to obtain
\begin{equation}
    \begin{split}
        & \log P \left( \bm{\theta} \vert \bm{\xo}, \bm{\yo} \right) \\
    & = - \frac{1}{2} \sum_i \frac{ w^2 \left( A \xo^{(i)} + B - \yo^{(i)} \right)^2 + {\sigma_x^{(i)}}^2 \left( A \mu + B - \yo^{(i)} \right)^2}{A^2 w^2 {\sigma_x^{(i)}}^2 +  {s_y^{(i)}}^2 \left( w^2 + {\sigma_x^{(i)}}^2 \right)} \\
    & - \frac{1}{2} \sum_i \frac{ {s_y^{(i)}}^2 \left( \xo^{(i)} - \mu \right)^2}{A^2 w^2 {\sigma_x^{(i)}}^2 +  {s_y^{(i)}}^2 \left( w^2 + {\sigma_x^{(i)}}^2 \right)} \\
    & - \frac{1}{2} \sum_i \log \left(A^2 w^2 {\sigma_x^{(i)}}^2 + {s_y^{(i)}}^2 \left( w^2 + {\sigma_x^{(i)}}^2 \right) \right)+ {\rm const}.
    \end{split}
\end{equation}

Again, we consider the case of equal errors for all data points to gain analytic insight into the bias of this method. As derived in Section~\ref{sec:app mnr}, the maximum likelihood values are
\begin{align}
    \hat{A} &=  \frac{\cov \left( \bm{\xo}, \bm{\yo} \right)}{\var \left( \bm{\xo} \right) - \sigma_x^2}, \\
    \hat{B} &= \langle \bm{\yo} \rangle - \hat{A} \langle \bm{\xo} \rangle \\
    \hat{s}^2 &= \var \left( \bm{\yo} \right) -
    \frac{\cov \left( \bm{\xo}, \bm{\yo} \right)^2}{\var \left( \bm{\xo} \right) - \sigma_x^2},
\end{align}
which are the same as the results for the uniform case, except that we replace $\var \left( \bm{\xo} \right)$ with $\var \left( \bm{\xo} \right) - \sigma_x^2$ in those expressions. As demonstrated in Section~\ref{sec:app mnr}, this is the correction required to remove the bias of the \unif{} method, and the maximum of the \mnr{} posterior occurs at the true parameter values irrespective of the underlying $\bm{\xt}$ distribution for large sample sizes. Much of the rest of the paper is devoted to showing that this remains true when the assumptions of constant uncertainties and large sample size are relaxed, and hence that \mnr{} is a robust solution to the regression problem.

It is possible for one Gaussian not to be sufficiently flexible to capture the $\bm{\xt}$ distribution. The simplest way to increase flexibility is to replace the Gaussian by a GMM, viz. a sum of $N_{\rm g}$ Gaussians each with their own weights, means and variances.
We will find in Sec.~\ref{sec:gmm} that a single Gaussian is typically sufficiently flexible to eliminate bias even in cases where the true distribution of $\bm{\xt}$ is highly non-Gaussian, and hence that adding more parameters to the prior may not be justified given the increased complexity. Nevertheless we will show cases where adding more Gaussians does reduce small residual biases, and provide this as an option in \textsc{roxy}.

\section{Numerical experiments}
\label{sec:mock}

Having established the basic properties of the three methods for integrating out $\bm{\xt}$, we now construct a systematic suite of numerical experiments to investigate their performance in full generality. This appears not to have been done before for any method, possibly because with less efficient sampling methods than Hamiltonian Monte Carlo it is computationally infeasible to sample the stochasticity of mock datasets thoroughly while spanning a high-dimensional parameter space. We achieve it by generating and fitting mock data sets with the following relevant properties:
\begin{itemize}
    \item True (generating) values of slope, intercept and intrinsic scatter $\tilde{A}$, $\tilde{B}$ and $\tilde{\sigma}_{\rm int}$.
    \item True (generating) distribution of $\bm{\xt}$, $\bm{\tilde{x}_{\rm t}}$.
    \item Number of datapoints $N$.
    \item Distribution of $x$ and $y$ uncertainties.
\end{itemize}
However, not all of these are independently relevant for the quality of the regression:
there are four degrees of freedom related to the translation and scaling invariance of the problem (where one puts the origin, and the units of $x$ and $y$). Fixing one of the parameters of the $\bm{\tilde{x}_{\rm t}}$ distribution and the true intercept, $\tilde{B}$, sets the origin, and the second parameter of the $\bm{\tilde{x}_{\rm t}}$ distribution and the mean $y$ error sets the units.
Thus we fix $\tilde{B}=1$ and $\sigma_y^{(i)} \hookleftarrow \mathcal{G}\left(2, 0.2\right)$. We take the mean $\sigma_x$, $\bar{\sigma}_x$, to be a free parameter, and set $\sigma_x^{(i)} \hookleftarrow \mathcal{G}\left(\bar{\sigma}_x, \bar{\sigma}_x/5\right)$, where $\mathcal{G}$ is a Gaussian distribution with a mean given by the first argument and standard deviation given by the second.

We consider three $\bm{\tilde{x}_{\rm t}}$ probability distributions: a uniform between 0 and 30, a triangular (or ramp) distribution rising linearly from 0 to 30, and an exponential peaking at 0 with a scale length of $\lambda_x$. The results are qualitatively similar in each case, with the exponential distribution, being the most non-Gaussian, providing the toughest test for \mnr. We therefore focus on this distribution for the remainder of the section.
The mock data is generated by first drawing $N$ uncorrelated $\bm{\tilde{x}_{\rm t}}$ samples, then calculating the true $y$ values as $\tilde{y}_\text{t}^{(i)} = \tilde{A} \tilde{x}_{\rm t}^{(i)} + \tilde{B}$. The observed values are then generating according to $\xo^{(i)} \hookleftarrow \mathcal{G}\left(\tilde{x}_{\rm t}^{(i)}, \sigma_x^{(i)}\right)$ and $\yo^{(i)} \hookleftarrow \mathcal{G}\left(\tilde{y}_\text{t}^{(i)}, \sqrt{{\sigma_y^{(i)}}^2 + \sint^2}\right)$.

In all cases we derive posteriors using the No U-Turns (NUTS) method of Hamiltonian Monte Carlo (HMC) as implemented in the \textsc{numpyro} sampler \citep{phan2019composable,bingham2019pyro}. We use 700 warmup and 5000 sampling steps and verify that the Gelman-Rubin statistic $r$ satisfies $r-1<10^{-3}$. We impose uniform improper priors on $A$, $B$ and $\sint$, such that $A$ and $B$ can take any value and $\sint$ is forced to be positive.
Similarly, in the case of \mnr, we place
improper uniform priors on $\mu$ and $w$, where $\mu$ can take any value and $w$ is positive.
These are the defaults in \textsc{roxy}. 

\subsection{1D variations}\label{sec:1D}

The remaining parameter space comprises $\tilde{A}$, $\tilde{\sigma}_{\rm int}$, $N$, $\bar{\sigma}_x$ and $\lambda_x$. As a first test, we fix four of these to fiducial values, as shown in Table~\ref{tab:params}, and vary the fifth within its full range using 20 equal steps.
For each value of this free parameter, and for each likelihood,
we generate 150 mock data sets differing only in the random seed used to generate the $\bm{\tilde{x}_{\rm t}}$, $\sigma_x$, $\sigma_y$, $\bm{\xo}$ and $\bm{\yo}$ values. For each one we calculate the bias in parameter $p$ as $(\bar{p}-\tilde{p})/\delta p$. For $p=\{A,B\}$ the overbar denotes the mean of the parameter's posterior and $\delta$ its standard deviation. For $p=\sint$, due to the lower bound of 0, they instead denote the mode and standard deviation of a normal fit that has been truncated at 0. We summarise the results over the 150 mock data sets by the median and 68\% confidence interval of this bias, plotting them against the parameter that is varying.

\begin{table}
    \centering
    \begin{tabular}{lccc}
        Property & Description & Fiducial value & Range \\
        \hline
         & $x$-range of data & [0,30] & --\\
        $\tilde{A}$ & True slope & 5 & [-30, 30]\\
        $\tilde{B}$ & True intercept & 1 & --\\
        $\tilde{\sigma}_{\rm int}$ & True intrinsic scatter & 2 & [0, 20]\\
        $N$ & Number of datapoints & 1000 & [10, 4000]\\
        $\bar{\sigma}_y$ & Mean $y$-error size & 2 & --\\
        $\bar{\sigma}_x$ & Mean $x$-error size & 1 & [0, 20]\\
        $\delta{\sigma}_y$ & St. dev. of $y$-error & 0.2 & --\\
        $\delta{\sigma}_x$ & St. dev. of $x$-error & $\bar{\sigma}_x/5$ & --\\
        $\lambda_x$ & Exponential scalelength & 8 & [1, 15]\\
        \hline
    \end{tabular}
    \caption{Properties of the mock data sets used in our numerical experiments. The fiducial value is used in cases where the parameter is fixed (see Sec.~\ref{sec:1D}). Parameters that are never varied have range quoted as ``--''.}
    \label{tab:params}
\end{table}

\begin{figure*}
  \centering
  \includegraphics[width=0.33\textwidth,height=0.25\textwidth]{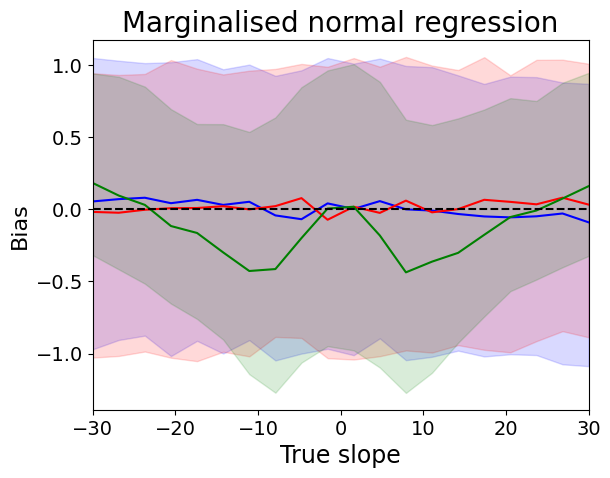}
  \includegraphics[width=0.33\textwidth,height=0.25\textwidth]{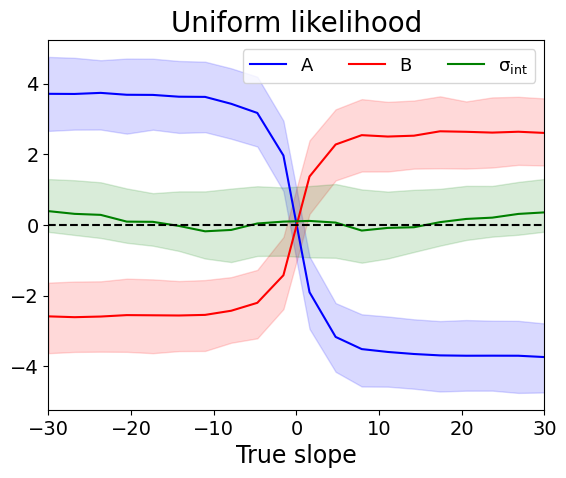}
  \includegraphics[width=0.33\textwidth,height=0.25\textwidth]{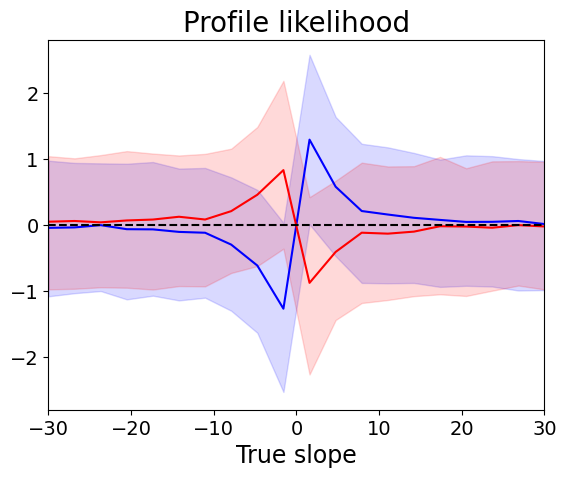}
  \includegraphics[width=0.33\textwidth,height=0.25\textwidth]{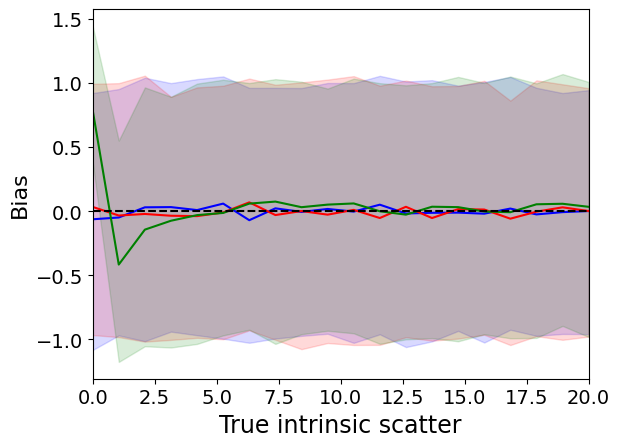}
  \includegraphics[width=0.33\textwidth,height=0.25\textwidth]{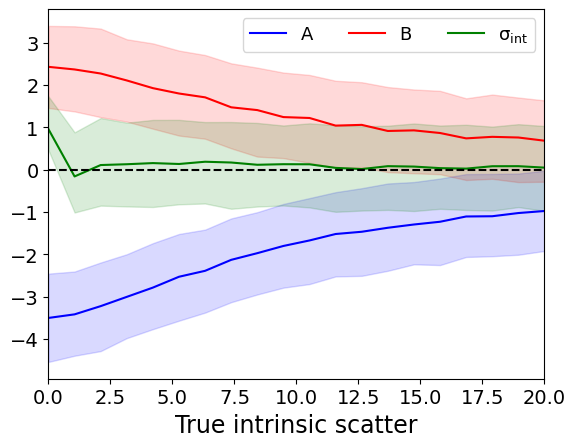}
  \includegraphics[width=0.33\textwidth,height=0.25\textwidth]{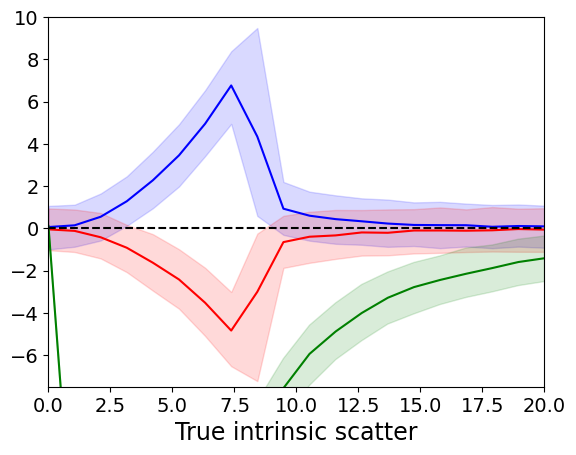}
  \includegraphics[width=0.33\textwidth,height=0.25\textwidth]{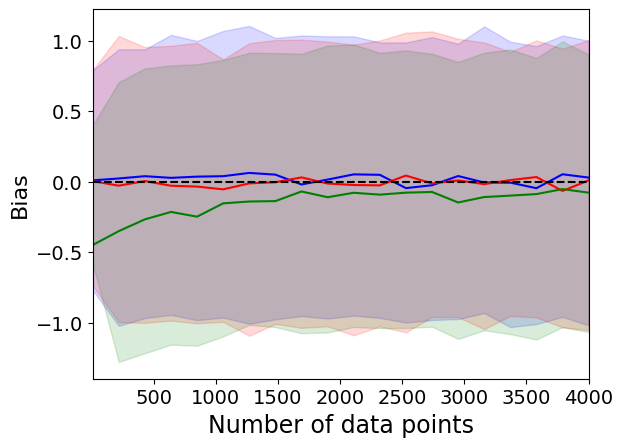}
  \includegraphics[width=0.33\textwidth,height=0.25\textwidth]{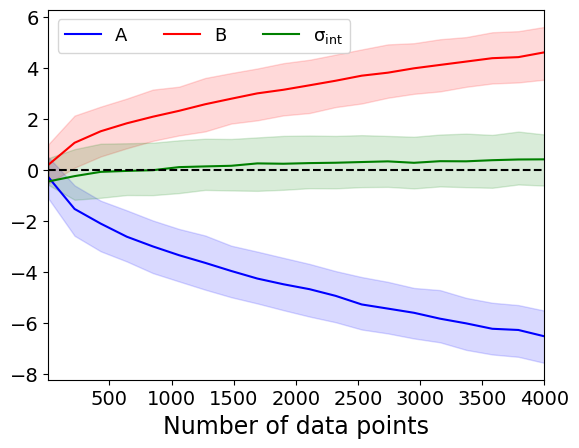}
  \includegraphics[width=0.33\textwidth,height=0.25\textwidth]{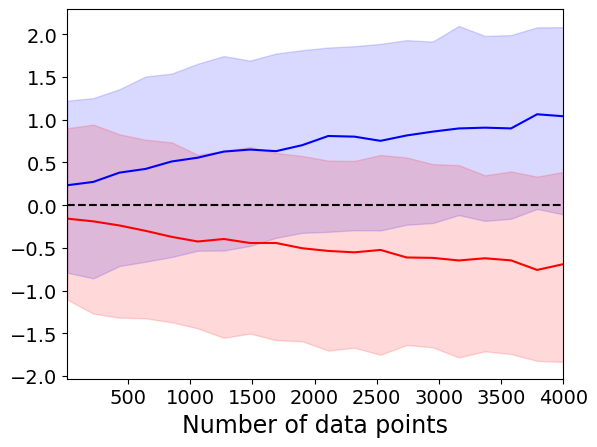}
  \includegraphics[width=0.33\textwidth,height=0.25\textwidth]{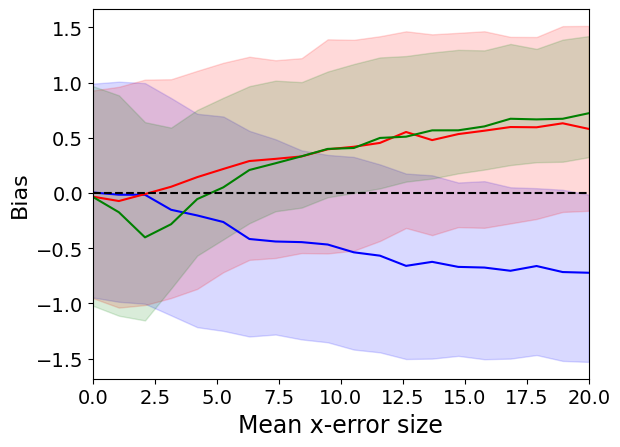}
  \includegraphics[width=0.33\textwidth,height=0.25\textwidth]{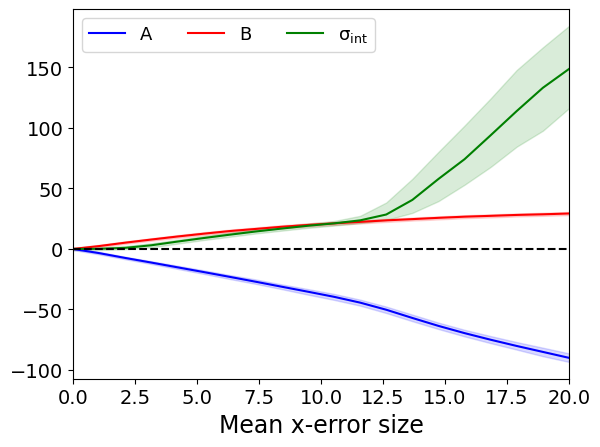}
  \includegraphics[width=0.33\textwidth,height=0.25\textwidth]{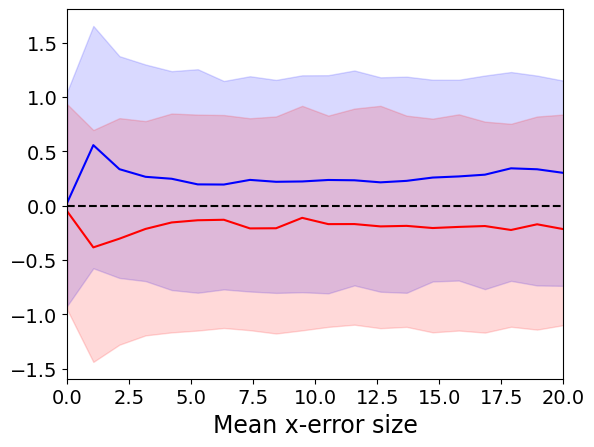}
  \includegraphics[width=0.33\textwidth,height=0.25\textwidth]{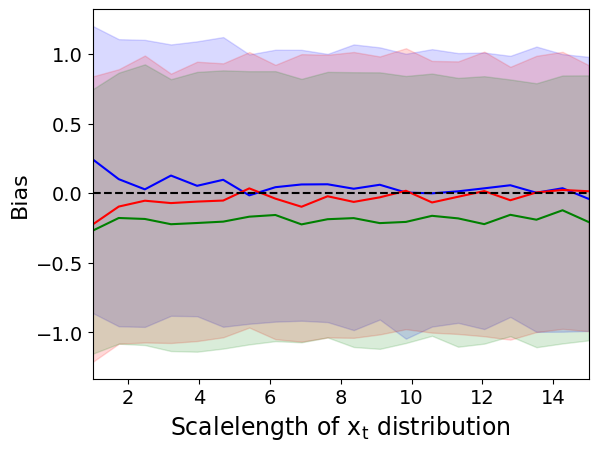}
  \includegraphics[width=0.33\textwidth,height=0.25\textwidth]{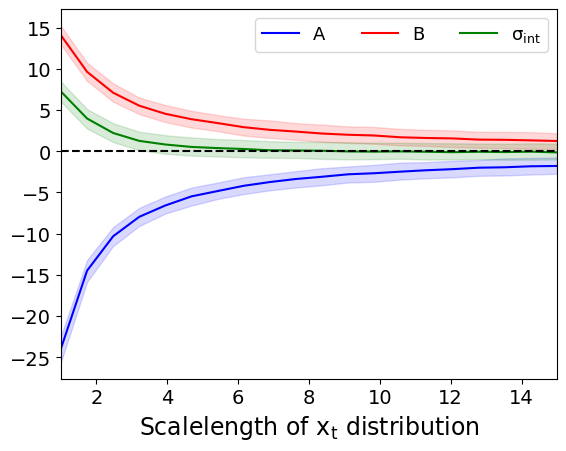}
  \includegraphics[width=0.33\textwidth,height=0.25\textwidth]{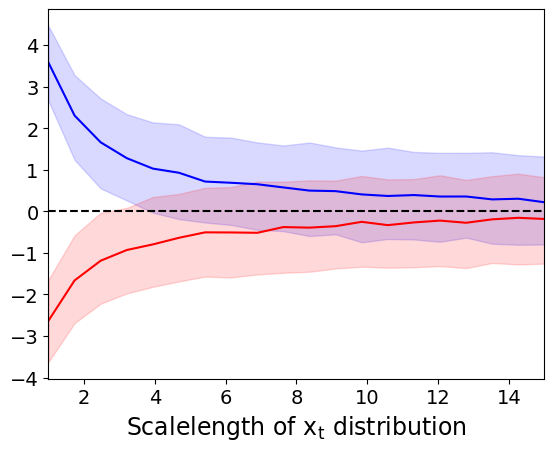}
  \caption{Distribution of biases incurred by the \mnr{} (left), \unif{} (centre) and \prof{} (right) methods in the slope (blue), intercept (red) and intrinsic scatter (green) of the straight line as various parameters of the problem are varied with the others held at their fiducial values (see Table.~\ref{tab:params}). In most cases \prof{} gives such strong negative bias in $\sint$ that the green curves lie far beyond the range of the plot. While the \unif{} and \prof{} methods are often wildly biased, \mnr{} has a sub-1$\sigma$ average bias in all cases, with the bias distribution typically conforming very closely to the expected standard normal. 
  }
  \label{fig:1D}
\end{figure*}

These results are shown in Fig.~\ref{fig:1D}. We see immediately that for \mnr{} the bias values are centred at approximately 0 in all cases, with a 68\% confidence bound extending to $\pm$1$\sigma$ bias as expected. The other methods regularly achieve large biases and are therefore highly unreliable. This is particularly severe in the \prof{} inference of $\sint$ which is pinned to 0 for $\tilde{\sigma}_{\rm int}\lesssim9$ for our fiducial parameter values; this produces such large biases that typically they cannot conveniently be shown on the plots.

\begin{figure}
  \centering
  \includegraphics[width=0.495\textwidth]{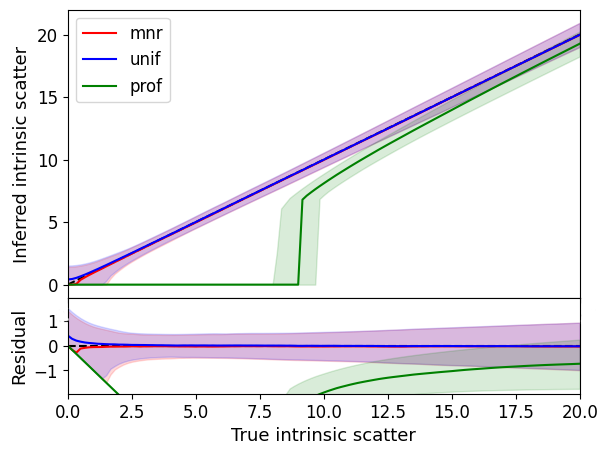}
  \caption{True intrinsic scatter vs that inferred by the \prof, \unif{} and \mnr{} methods with all other parameters at their fiducial values (see Table~\ref{tab:params}). For each $\tilde{\sigma}_\text{int}$ value we generate 300 mock data sets and show the median and 95\% range of the distribution of maximum-likelihood values, along with the residuals with respect to the truth in the lower panel.
  }
  \label{fig:sig_comp}
\end{figure}

To supplement Fig.~\ref{fig:1D}, we show in Fig.~\ref{fig:sig_comp} the true vs inferred $\sint$ for the three methods at the fiducial values of the other parameters. We calculate here only the maximum-likelihood values (i.e. without posterior uncertainties), showing the median and 95\% spread from 300 random mock datasets at each $\tilde{\sigma}_\text{int}$ value for each method along with the residuals with respect to the truth for the \unif{} and \mnr{} methods in the lower panel. This shows that while the \unif{} and \mnr{} methods recover approximately correct $\sint$ from small to large values, the profile likelihood returns $\sint=0$ until $\tilde{\sigma}_{\rm int}$ is several times larger than $\sigma_x$ and $\sigma_y$, beyond which it is slightly biased low. We have already observed this behaviour in Section~\ref{sec:prof}.

\subsection{5D variations}\label{sec:5D}

By fixing all but one parameter, this test does not provide a thorough exploration of the parameter space of possible data sets. While we readily see that the \prof{} and \unif{} methods are highly biased, there may be datasets where even \mnr{} is unreliable.
We therefore now consider the full 5D grid in $\tilde{A}$, $\tilde{\sigma}_{\rm int}$, $N$, $\bar{\sigma}_x$ and $\lambda_x$, sampled uniformly (logarithmically in $N$) between the limits [-30,30], [0,20], [10,4000], [0,20] and [1,15] respectively with 5 grid points along each dimension. At each grid point we generate 150 mock datasets with the corresponding parameters and fit them using \mnr, calculating in each case the bias in $A$, $B$ and $\sint$ as defined above.
We then locate the points in the 5D space where the average bias in $A$, $B$ or $\sint$ is greatest in either direction, and at these points we calculate the distribution of bias values in the three parameters over the mock data sets.

\begin{figure}
  \centering
  \includegraphics[width=0.495\textwidth]{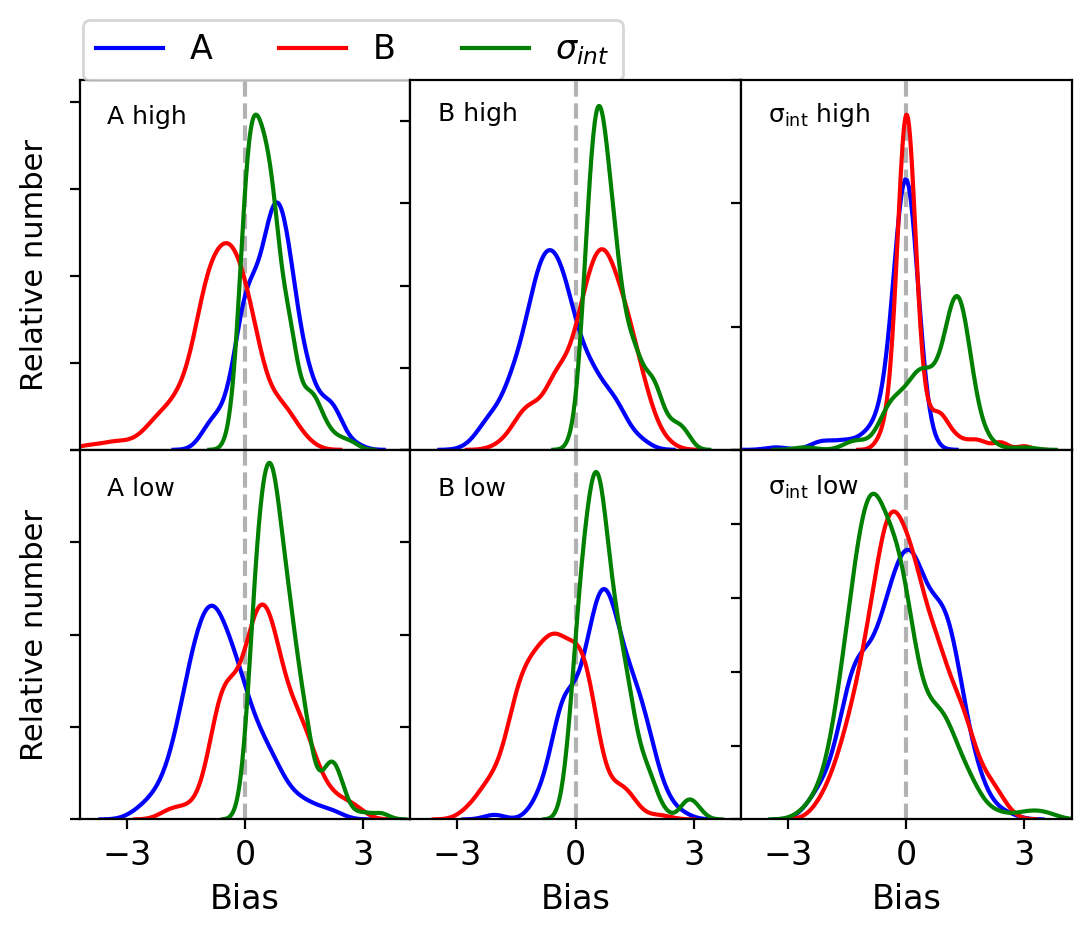}
  \caption{Distribution of biases incurred by \mnr{} at the most biased points in our 5-parameter grid search. Separate panels show the results when the slope, intercept and normalisation of the fit are separately maximally biased high and low, and in each case we show the bias in all three parameters. The average bias never exceeds $\sim$1$\sigma$, and for any mock dataset's random stochasticity it does not exceed $\sim$3$\sigma$.
  }
  \label{fig:5D}
\end{figure}

We find the points in the [$\tilde{A}$, $\tilde{\sigma}_{\rm int}$, $N$, $\bar{\sigma}_x$, $\lambda_x$] space which produce the largest biases are
\begin{itemize}
\item A most biased high: [-15, 894, 10, 20, 15],
\item A most biased low: [15, 4000, 0, 15, 15],
\item B most biased high: [15, 4000, 0, 20, 15],
\item B most biased low: [-15, 4000, 5, 20, 15],
\item $\sint$ most biased high: [15, 10, 0, 20, 15],
\item $\sint$ most biased low: [15, 4000, 10, 5, 15].
\end{itemize}
The distributions of biases in each parameter at these points are shown in Fig.~\ref{fig:5D}. We see that the average bias of \mnr{} never exceeds $\sim$1$\sigma$, showing it to be fully reliable across the parameter space. These maximum biases are typically obtained for large $x$-errors, a nearly flat $\bm{\xt}$ distribution and very large sample size.
We have repeated this calculation for the \prof{} and \unif{} methods (not shown), where we find enormous maximum biases: $\sim$(10, 10, 750)$\sigma$ for ($A$, $B$, $\sint$) in \prof{} and $\sim$(200, 1250 and 500)$\sigma$ for \unif.

\subsection{Gaussian mixture model}\label{sec:gmm}

Finally, we consider whether the use of additional Gaussians in the $\bm{\xt}$ prior can alleviate the small biases seen on occasion for \mnr. An important consideration here is the choice of priors assigned to the hyperparameters describing the Gaussians. We consider two possibilities: 1) uniform priors 
(although requiring that the means of the Gaussians are ordered to break the degeneracies between the different Gaussians during sampling),
paralleling the case of \mnr, and 2) a hierarchical set of priors as described in~\citet{Roeder_Wasserman_1997,Carroll_1999,Kelly_2007} (see also~\citealt{Escobar_West}). This introduces hyper-hyperparameters coupling the properties of the Gaussians, favouring similar means and widths as \citeauthor{Kelly_2007} suggests to be a desirable property.

In the ``hierarchical'' model, a Gaussian prior is placed on the mean of each Gaussian in the mixture, with some mean $\mu_\ast$ and variance $u_\ast^2$. Additionally, a scaled inverse $\chi^2$ prior is adopted on the variance of each Gaussian, with one degree of freedom and scale $w_\ast^2$. This introduces three additional parameters to the inference: $\mu_\ast$, $u_\ast^2$ and $w_\ast^2$. A uniform prior is placed on $\mu_\ast$ and $w_\ast^2$, and a scaled inverse $\chi^2$ prior is chosen for $u_\ast^2$, with one degree of freedom and scale parameter $w_\ast^2$.
We find the two sets of priors to give very similar results, and use the hierarchical set fiducially as it is more common in the literature.

How many Gaussians should one use? In principle one can take the number of Gaussians $N_{\rm g}$ as a free parameter and marginalise over it as part of the inference~\citep{Richardson_2002,Richardson_Green}. Since the number of parameters in the inference is a function of $N_{\rm g}$ 
this requires trans-dimensional inference methods such as reversible jump MCMC~\citep{RJ}. An alternative is to use the Bayesian evidence to determine the value of $N_{\rm g}$ that optimally trades off the gain in likelihood when using more Gaussians with the increase in the number of free parameters. This is however difficult to compute, so one may wish to approximate it through the Bayesian Information Criterion (BIC). This is defined as \cite{BIC}
\begin{equation}
\text{BIC} = k\ln(N) - 2\ln(\hat{\mathcal{L}})
\end{equation}
where $k$ is the number of free parameters of the model.
The BIC is the limit of the evidence when the posterior is modelled as a multi-dimensional Gaussian around the maximum a posteriori point and the number of data points is much larger than the number of free parameters.
We implement this method in our code, providing a simple way of determining the optimum model. Note however that using more Gaussians does not necessarily lead to a reduction in bias~\citep{Carroll_1999}, and if there are too few datapoints the inference of a large number of Gaussian hyperparameters may not be feasible. The safest choice is therefore to perform a sensitivity analysis, examining the results for a range of different $N_{\rm g}$ (conservatively 1-10). The constraints are robust if they are stable across multiple $N_{\rm g}$ values.

\begin{figure}
  \centering
  \includegraphics[width=0.49\textwidth]{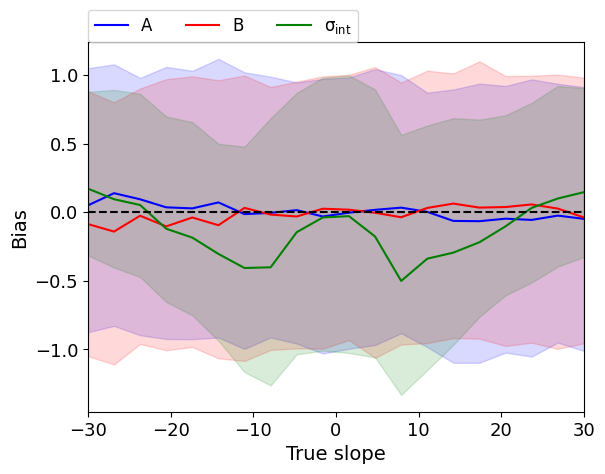}
  \includegraphics[width=0.49\textwidth]{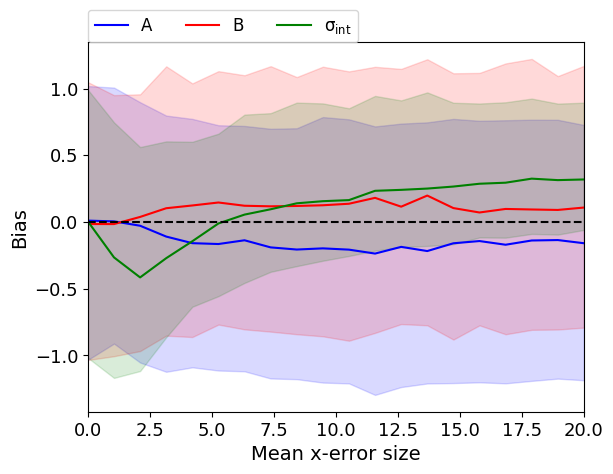}
  \caption{Distribution of biases incurred during variation of the true slope (top) and average $x$-error size (bottom; see Fig.~\ref{fig:1D}) when describing the $\bm{\xt}$ distribution using two Gaussians. The bias is reduced somewhat in the latter case, but not the former.}
  \label{fig:1D_2Gauss}
\end{figure}

To begin investigating the impact of $N_{\rm g}$, we repeat the procedure used in Fig.~\ref{fig:1D} for varying slope (first row) and mean $x$-error size (fourth row) but using two Gaussians instead of one.
The results are shown in Fig.~\ref{fig:1D_2Gauss}, where we see that the bias in the first case is unaffected while in the latter it is somewhat reduced.

\begin{figure*}
  \centering
  \includegraphics[width=\textwidth]{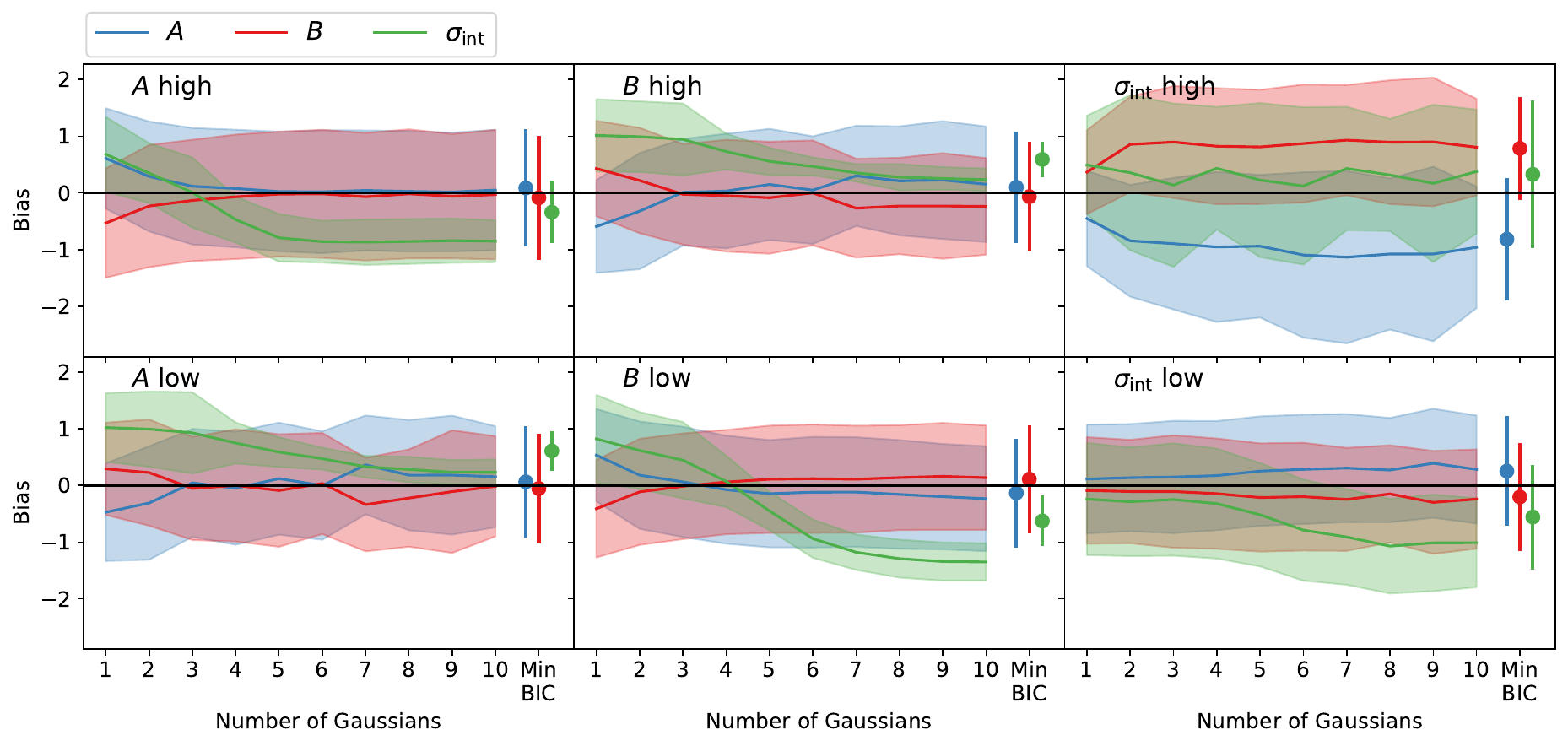}
  \caption{Distribution of biases across 150 mock data sets as a function of the number of Gaussians used in the GMM for the same parameters as Fig.~\ref{fig:5D}. We also show the distribution of biases for the number of Gaussians that minimises the BIC, which can be different for each dataset. The bars and bands show the $1\sigma$ deviation among the mock data sets. Any improvement by increasing the number of Gaussians or choosing the best number according to the BIC is marginal, suggesting that one Gaussian cannot generically be improved upon.
  }
  \label{fig:bias_ngauss}
\end{figure*}

\begin{figure}
  \centering
  \includegraphics[width=\columnwidth]{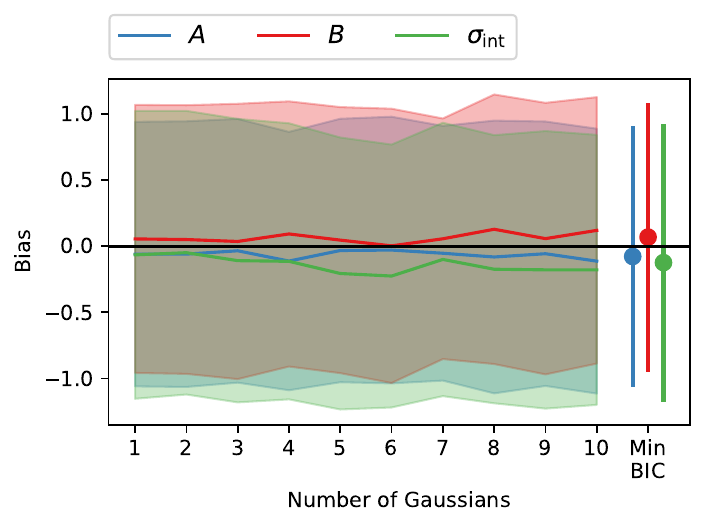}
  \caption{Same as Fig.~\ref{fig:bias_ngauss}, but using the fiducial parameters given in Table~\ref{tab:params}. The bias in each parameter is consistent with zero in all cases, and the very small mean biases in the case of one Gaussian are not reduced by using more.}
  \label{fig:bias_ngauss_fiducial}
\end{figure}

We now consider the cases which were most biased when only using one Gaussian (Fig.~\ref{fig:5D}). Here we rerun the inference for each dataset using a GMM containing between 1 and 10 Gaussians 
(with 150 mock datasets as before),
and plot the bias as a function of $N_{\rm g}$ in Fig.~\ref{fig:bias_ngauss}. For each dataset, we compute the BIC for each number of Gaussians, and show as a separate case the bias distribution when choosing the number of Gaussians that minimises this. The distribution of this number for each run is shown
in Table~\ref{tab:bias_ngauss}. We observe that the bias distribution often changes very little as one varies the number of Gaussians, such that there is not a significant, systematic difference between choosing just one Gaussian, the number that minimises the BIC or any other. So, although it may seem preferable to
choose more than one Gaussian to find a better approximation to the distribution of $\bm{\xt}$, this does not appear to be necessary to improve the inference on $A$, $B$ or $\sint$. Indeed, in some cases the bias is \emph{increased} by increasing $N_{\rm g}$. Also recall that the parameters used in Fig.~\ref{fig:bias_ngauss} are those which are most biased when using one Gaussian, such that one may expect the biggest improvement by increasing this number. In Fig.~\ref{fig:bias_ngauss_fiducial} we repeat this analysis using the fiducial parameters (for which one Gaussian already works near-perfectly) and see that the effect of increasing the number of Gaussians is even smaller.

\begin{table}
	\centering
	\begin{tabular}{lcccccccccc}
	\multirow{2}{*}{Run} & \multicolumn{10}{c}{Number of Gaussians that minimises BIC} \\
	& 1 & 2 & 3 & 4 & 5 & 6 & 7 & 8 & 9 & 10 \\
		\hline
		$A$ high &  &  & 49 & 89 & 12 &  &  &  &  &  \\
		$A$ low &  & 1 & 7 & 31 & 73 & 29 & 5 & 4 &  &  \\
		$B$ high & 1 &  & 6 & 35 & 66 & 30 & 10 &  & 1 & 1 \\
		$B$ low &  &  &  & 6 & 85 & 55 & 4 &  &  &  \\
		$\sigma_{\rm int}$ high & 30 & 66 & 32 & 12 & 5 & 2 & 1 &  & 1 & 1 \\
		$\sigma_{\rm int}$ low &  &  &  & 13 & 105 & 32 &  &  &  &  \\
		Fiducial &  &  & 49 & 83 & 16 & 1 &  & 1 &  &  \\
	\end{tabular}
	\caption{Distribution of the number of Gaussians that minimises the BIC for each of the runs considered in Fig.~\ref{fig:bias_ngauss}
 (out of 150 mock data sets).
 For these runs, the mode of the distribution is always $\leq5$, suggesting that a maximum $N_{\rm g}$ of 10 is sufficient to identify the minimum BIC in all cases.}
        \label{tab:bias_ngauss}
\end{table}

\section{Real-world test case: The masses of galaxy clusters}\label{sec:applications}

While the experiments of Section~\ref{sec:mock} show the possible differences between the \unif, \prof{} and \mnr{} results, many of the parameter choices were deliberately rather extreme to stress-test the methods.
We now apply the methods to astrophysical data 
to illustrate the importance of the method in a real-world setup. 

Galaxy clusters, tracing the densest regions of the universe, are a fundamental tool in cosmology. However, their most basic property, mass, is not directly observable, and must therefore be estimated by means of an observable proxy. This may be the cluster's ``richness'' (the number of galaxies it contains), the luminosity of X-rays emitted by hot gas in the intra-cluster medium, the velocity dispersion of galaxies in the cluster or the Sunyaev-Zel'dovich (SZ) decrement \citep{Sunyaev_1972} when observing the cosmic microwave background through the cluster.
Here we directly compare the mass estimates using the latter two methods. We consider clusters from the second \textit{Planck} \citep{Planck_2014} catalogue of SZ clusters \citep{Planck_clusters_2016} which have optical counterparts in the 128- MULTIPLE-16/15B (LP15) observational program \citep{Aguado-Barahona_2019,Streblyanska_2019} and Sloan Digital Sky Survey \citep{SDSS_2000}, as recorded in \citet{Aguado-Barahona_2022}.

As in \citet{Aguado-Barahona_2022}, we assume that the SZ-determined mass, $M_{500}^{\rm SZ}$, is related to that from dynamical mass estimates, $M_{500}^{\rm dyn}$, according to
\begin{equation}
    \label{eq:cluster mass relation}
    \log \left( \frac{M_{500}^{\rm SZ}}{6 \times 10^{14} {\rm \, M_{\odot}}} \right) = 
    \alpha \log \left( \frac{M_{500}^{\rm dyn}}{6 \times 10^{14} {\rm \, M_{\odot}}} \right)
    + \log \left( 1 - b \right),
\end{equation}
where $\alpha$ and $b$ are free parameters describing the slope and intercept of the relation. 
$1-b$ is interpreted as the ratio of biases between the two measured masses and the true masses of the objects, which would be unity if both methods accurately determined the true mass (or had the same bias).

To fit this relation, \citeauthor{Aguado-Barahona_2022} consider a number of methods: orthogonal distance regression \citep{ODR}, the Nukers method \citep{Tremaine_2002}, the method we dub \unif{} (Sec.~\ref{sec:unif}), the bivariate correlated errors and intrinsic scatter method \citep{Akritas_1996}, and the Gaussian mixture model method of \citet{Kelly_2007} (which they call Complete Maximum Likelihood Estimation, CMLE). They find that all of these methods give biased results when applied to mock data, but that the Nukers method gives the most reliable bias, and is thus easiest to correct. In Section~\ref{sec:mock} we demonstrated that \mnr{} robustly gives unbiased results for a range of settings and distributions of $\bm{\xt}$, and the properties of this dataset lie well within the 5D parameter space we explore.
As before, we fit the \prof, \unif{} and \mnr{} methods using uniform priors on the slope and intercept of Eq.~\ref{eq:cluster mass relation}, the latter of which we subsequently convert to $1-b$ through exponentiation.

\begin{figure*}
\centering
\begin{minipage}[b]{.49\textwidth}
 \centering
    \includegraphics[width=\columnwidth]{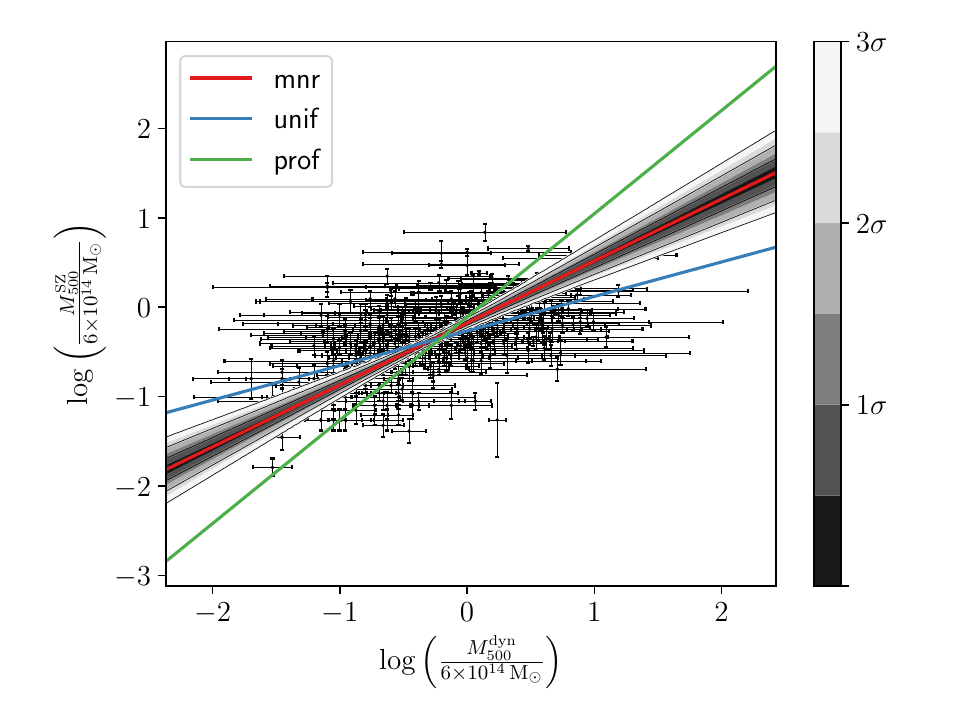}
    \caption{SZ-inferred cluster mass, $M_{500}^{\rm SZ}$, versus dynamical mass, $M_{500}^{\rm dyn}$, for \textit{Planck} SZ clusters with optical counterparts, as given in \citet{Aguado-Barahona_2022} (black points with errorbars). We plot the posterior predictive distribution of the \mnr-inferred relation (Eq.~\ref{eq:cluster mass relation}) in greyscale, including the confidence range up to
3$\sigma$, and the posterior mean predictions for all three methods are plotted as coloured lines.}
    \label{fig:cluster predictive}
\end{minipage}
\begin{minipage}[b]{.49\textwidth}
\centering
    \includegraphics[width=\columnwidth]{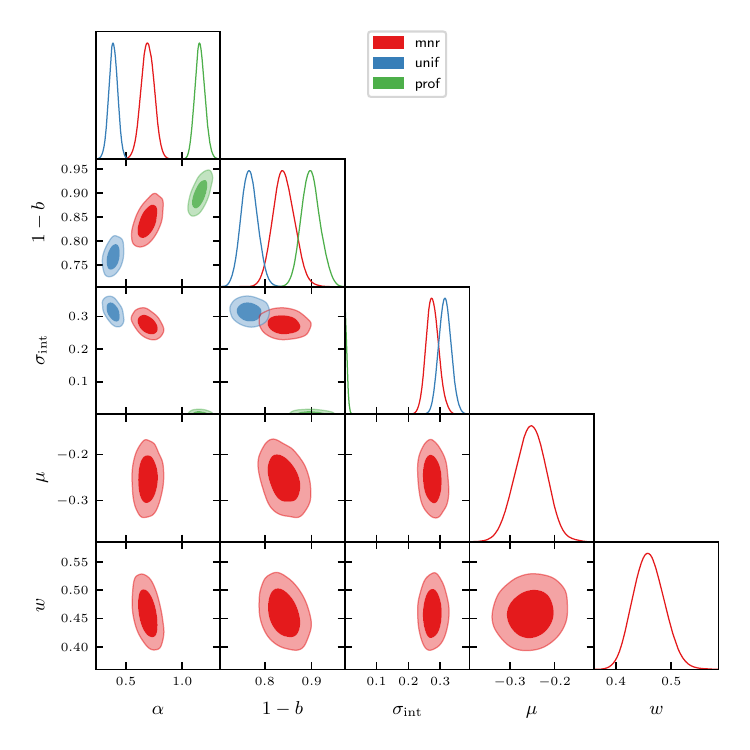}
    \caption{Corner plot of the slope, normalisation and intrinsic scatter of the relation between SZ-inferred cluster masses and dynamical masses, as given in Eq.~\ref{eq:cluster mass relation}, alongside the parameters of the Gaussian hyperprior for \mnr. In this case the $x$-errors result in significant differences between the methods.
}
    \label{fig:cluster corner}
\end{minipage}
\end{figure*}

The results are shown in Fig.~\ref{fig:cluster predictive}. We see that the $x$-uncertainties for the relation are sizeable, with average values for $\log M_{500}^{\rm dyn}$ and $\log M_{500}^{\rm SZ}$ of 0.44 and 0.10, respectively. The large number of data points means that we have a reasonable constraint on the function within the range of the data, as shown by the posterior predictive distribution for \mnr. However, the sizes of the uncertainties result in the three methods giving very different results for the parameter values. We plot the two-dimensional posteriors for $\alpha$, $1-b$ and $\sigma_{\rm int}$ (and $\mu$ and $w$ in the case of \mnr) in Fig.~\ref{fig:cluster corner}. While $\sint$ is 0 for the \prof{} method and compatible between \unif{} and \mnr{} (cf. Fig.~\ref{fig:sig_comp}),
the values of $\alpha$ and $1-b$ are in significant disagreement between the three methods:
\begin{align}
    \mnr: \quad \alpha &= 0.70 \pm 0.06, \quad 1 - b = 0.84 \pm 0.02, \\
    \unif: \quad \alpha &= 0.39 \pm 0.04, \quad 1 - b = 0.77 \pm 0.02, \\
    \prof: \quad \alpha &= 1.16^{+0.05}_{-0.04}, \quad \ \ \; 1 - b = 0.90 \pm 0.02.
\end{align}
The inferred value from the \prof{} method is compatible with that reported by \citeauthor{Aguado-Barahona_2022}, since the ``Nukers'' method used in that paper is essentially the same as \prof{} given that the inferred intrinsic scatter is consistent with zero.
The small inferred $\alpha$ and the large inferred $\sint$ for the \unif{} method is exactly the trend expected given the bias of this method discussed in Section~\ref{sec:unif} and computed in Section~\ref{sec:app uniform}. We thus see that in this case it is crucial to use the principled and unbiased \mnr{} method rather than any of the other $\bm{\xt}$ marginalisation methods or ad hoc algorithms proposed in the literature.

Our best-fit value of the slope is very similar to that quoted by \citeauthor{Aguado-Barahona_2022} for the CMLE method implemented in the Python port of~\citeauthor{Kelly_2007}'s IDL code. They find this slope to be too low by $\sim$20\% in mock data, which, given the posterior uncertainties, amounts to a $\sim$3$\sigma$ bias. This is unexpected as CMLE is approximately equivalent to our GMM method, which we have shown to give very similar results to \mnr. To investigate this in the context of our bias results we generate mock data identical to the cluster data except that the $\yt$ values are generated according to the best-fit \mnr{} line, and with 
five models for the unknown $\bm{\xt}$ distribution. These are:
\begin{itemize}
\item $\xt^{(i)}$ = $\xo^{(i)}$ (then scattered by $\sigma_x^{(i)}$ to produce a new $\bm{\xo}$ in each mock dataset);
\item $\bm{\xt}$ set to their maximum \textit{a posteriori} values given the \mnr{} fit to the data;
\item $\xt^{(i)} \hookleftarrow \mathcal{G}(\mu, w)$ where $\mu$ and $w$ are the maximum \textit{a posteriori} mean and width of the $\bm{\xt}$ hyperprior from the \mnr{} fit;
\item $\xt^{(i)} \hookleftarrow \mathcal{U}(-0.8,0.6)$ (relatively narrow range at the centre of the dataset);
\item $\xt^{(i)} \hookleftarrow \mathcal{U}(\text{min}(\bm{\xo}), \text{max}(\bm{\xo}))$ (wide range).
\end{itemize}
For all choices except the second we find the distributions of bias values in each of 
the slope, intercept and intrinsic scatter
over many mock datasets to closely approximate a standard normal, as would be expected if the method is performing accurately.
In the second model, however, we find 
the slope
to be biased low and 
the intercept
high by around 1$\sigma$ on average. The tail of the bias distributions extend to $\sim \pm$3$\sigma$ in this case, so it is unlikely but possible that in a given mock data realisation \mnr{} could be biased by the amount attributed by 
\citeauthor{Aguado-Barahona_2022} to CMLE, although across the entire population the bias is no worse than is seen in Fig.~\ref{fig:5D} and therefore not a cause for serious concern. Alternatively it may be that the bias testing method of \citeauthor{Aguado-Barahona_2022} is itself inaccurate.

Upcoming work will rectify the biases caused by regression method in a slew of astrophysical and cosmological fits (von Hausegger et al. 2024, in prep).

\section{Extension to nonlinear functions and non-diagonal covariance}
\label{sec:nonlinear}

Until now we have supposed that the $x$ and $y$ errors are completely independent---there was no covariance among the observed $x$ values, among the observed $y$ values, or between $\bm{\xo}$ and $\bm{\yo}$---and we focused on linear regression problems. 
The \unif, \prof{}, \mnr{}
and GMM
likelihoods that we consider are readily generalisable to arbitrary covariance matrices and functional forms of the fit, provided the $x$-uncertainties are small compared to the curvature scale of the function.

We now assume that 
the true values of $\bm{y}_{\rm t} = (y_{\rm t}^{(1)}, \ldots , y_{\rm t}^{(N)})$ are related by some deterministic function to the true values of $\bm{x}_{\rm t} = (x_{\rm t}^{(1)}, \ldots , x_{\rm t}^{(N)})$:
\begin{equation}
	\label{eq:general function}
	\bm{y}_{\rm t} = \bm{f} \left( \bm{x}_{\rm t}, \bm{\theta}\right),
\end{equation}
where $\bm{\theta}$ is the parameter vector we wish to infer. Note that throughout this section we do not assume that the value of $\yt^{(i)}$ only depends on $\xt^{(i)}$ and $\bm{\theta}$, and thus the results hold for arbitrary functions which can combine different $x$ values in the calculation of any given $y$ value.
This generalises Eq.~\ref{eq:likelihood} to 
\begin{equation}
    \label{eq:general likelihood}
    \begin{split}
        \log \mathcal{L} &= - \frac{1}{2} \log \det \left( 2 \pi \mathsf{\Sigma}\right)  \\
   & - \frac{1}{2} \left( \left(\bm{\xt} - \bm{\xo}\right)^{\rm T},  \left(\bm{f}(\bm{\xt}) - \bm{\yo}\right)^{\rm T} \right)\mathsf{\Sigma}^{-1} 
    \begin{pmatrix}
           \bm{\xt} - \bm{\xo} \\
           \bm{f}(\bm{\xt}) - \bm{\yo} \\
     \end{pmatrix},
    \end{split}
\end{equation}
where we have dropped the dependence of $\bm{f}$ on $\bm{\theta}$ for convenience. Here $\mathsf{\Sigma}$ is the covariance matrix of the data, which we define as
\begin{equation}
    \mathsf{\Sigma} \equiv
    \begin{pmatrix}
        \mathsf{\Sigma}_{xx} & \mathsf{\Sigma}_{xy}\\
        \mathsf{\Sigma}_{yx} & \mathsf{\Sigma}_{yy}
    \end{pmatrix}
    \equiv
    \begin{pmatrix}
        \mathsf{\alpha} & \mathsf{\beta} \\
        \mathsf{\beta}^{\rm T} & \mathsf{\gamma}
    \end{pmatrix}^{-1}.
\end{equation}
If one wishes to include intrinsic scatter in the relation, one simply needs to add $\sigma_{\rm int}^2$ to each of the diagonal elements of $\mathsf{\Sigma}_{yy}$, and this can be inferred alongside $\bm{\theta}$.

We now assume that we can approximate the function $f$ as linear across the range spanned by each $x$-uncertainty. We therefore write
\begin{equation}
    \label{eq:Taylor series}
    \bm{f} \left( \bm{\xt} \right) \approx \bm{f} \left( \bm{\xo} \right) + \mathsf{G} \left( \bm{\xt} - \bm{\xo} \right),
\end{equation}
where
\begin{equation}
    \label{eq:G definition}
    \mathsf{G}_{ij} \equiv \frac{\partial}{\partial x^{(j)}} \left[ \bm{f} (\bm{x}) \right]_i \vert_{\bm{x} = \bm{\xo}}.
\end{equation}

Note that it may be possible to reparametrise a function that appears highly nonlinear to make it near-linear.
This is the case for example with power-law fits such as Eq.~\ref{eq:cluster mass relation}. 
Unlike the straight line case, whenever a nonlinear function is fitted Eq.~\ref{eq:Taylor series} causes some error, which the user is reminded to be mindful of. 
The approximation breaks down completely at points where the function's second derivative times $x$-error size exceeds the first derivative.

\subsection{Infinite uniform marginalisation}

Beginning with the case of infinite uniform marginalisation, we must exponentiate Eq.~\ref{eq:general likelihood} and substitute in Eq.~\ref{eq:Taylor series}. We see that to integrate this function over $\bm{\xt}$ and thus obtain the marginalised uniform likelihood, we have to perform a Gaussian integral
\begin{equation}
    \begin{split}
         P_{\rm U} &\approx \frac{1}{\sqrt{\det \left( 2 \pi \mathsf{\Sigma} \right)}}
            \exp \left[ - \frac{1}{2} \left(\bm{f} \left( \bm{\xo} \right) - \bm{\yo} \right)^{\rm T} \mathsf{\gamma} \left(\bm{f} \left( \bm{\xo} \right) - \bm{\yo} \right)\right] \\
            & \times \int {\rm d} \bm{z}
        \exp \left[ - \frac{1}{2}  \bm{z}^{\rm T} \mathsf{C} \bm{z} - \left(\bm{f} \left( \bm{\xo} \right) - \bm{\yo} \right)^{\rm T} \left( \mathsf{\beta}^{\rm T} + \mathsf{\gamma} \mathsf{G} \right) \bm{z} \right],
    \end{split}
\end{equation}
where 
\begin{equation}
    \mathsf{C} \equiv \mathsf{\alpha} + \mathsf{\beta} \mathsf{G} + \mathsf{G}^{\rm T} \mathsf{\beta}^{\rm T} + \mathsf{G}^{\rm T} \mathsf{\gamma} \mathsf{G}.
\end{equation}
Performing this integral, one obtains
\begin{equation}
    \label{eq:general uniform}
    \begin{split}
        \log P_{\rm U} \approx & - \frac{1}{2} \log \det \left( 2 \pi \mathsf{D}\right)  \\
        & - \frac{1}{2} \left(\bm{f} \left( \bm{\xo} \right) - \bm{\yo} \right)^{\rm T} \mathsf{D}^{-1} \left(\bm{f} \left( \bm{\xo} \right) - \bm{\yo} \right),
    \end{split}
\end{equation}
where
\begin{equation}
    \mathsf{D}^{-1} \equiv \mathsf{\gamma} - \left( \mathsf{\beta}^{\rm T} + \mathsf{\gamma} \mathsf{G} \right) \mathsf{C}^{-1} \left( \mathsf{\beta} + \mathsf{G}^{\rm T} \mathsf{\gamma} \right).
\end{equation}
To obtain an expression for $\mathsf{D}$, we use the Woodbury matrix identity
\begin{equation}
    \left( \mathsf{A} + \mathsf{U} \mathsf{C} \mathsf{V} \right)^{-1}
    = 
    \mathsf{A}^{-1} - \mathsf{A}^{-1} \mathsf{U} \left( \mathsf{C}^{-1} + \mathsf{V} \mathsf{A}^{-1} \mathsf{U} \right)^{-1} \mathsf{V} \mathsf{A}^{-1},
\end{equation}
to give
\begin{equation}
    \begin{split}
        \mathsf{D} =& 
    \mathsf{\gamma}^{-1} + \mathsf{\gamma}^{-1} \mathsf{\beta}^{\rm T} \left( \mathsf{\alpha} - \mathsf{\beta}\mathsf{\gamma}^{-1}\mathsf{\beta}^{\rm T} \right)^{-1} \mathsf{\beta}\mathsf{\gamma}^{-1}, \\
    &+ \mathsf{G} \left( \mathsf{\alpha} - \mathsf{\beta}\mathsf{\gamma}^{-1}\mathsf{\beta}^{\rm T} \right)^{-1} \mathsf{G}^{\rm T}
    + \mathsf{\gamma}^{-1} \mathsf{\beta}^{\rm T} \left( \mathsf{\alpha} - \mathsf{\beta}\mathsf{\gamma}^{-1}\mathsf{\beta}^{\rm T} \right)^{-1}  \mathsf{G}^{\rm T}, \\
    & + \mathsf{G} \left( \mathsf{\alpha} - \mathsf{\beta}\mathsf{\gamma}^{-1}\mathsf{\beta}^{\rm T} \right)^{-1} \mathsf{\beta} \mathsf{\gamma}^{-1}.
    \end{split}
\end{equation}
When one utilises the block-wise inversion formula for matrices, one can identify these terms with the elements of the covariance matrix used in Eq.~\ref{eq:general likelihood}, i.e.
\begin{align}
    \mathsf{\Sigma}_{yy} &= \mathsf{\gamma}^{-1} + \mathsf{\gamma}^{-1} \mathsf{\beta}^{\rm T} \left( \mathsf{\alpha} - \mathsf{\beta}\mathsf{\gamma}^{-1}\mathsf{\beta}^{\rm T} \right)^{-1} \mathsf{\beta}\mathsf{\gamma}^{-1} \\
    \mathsf{\Sigma}_{xx} &= \left( \mathsf{\alpha} - \mathsf{\beta}\mathsf{\gamma}^{-1}\mathsf{\beta}^{\rm T} \right)^{-1} \\
    \mathsf{\Sigma}_{xy} &= - \left( \mathsf{\alpha} - \mathsf{\beta}\mathsf{\gamma}^{-1}\mathsf{\beta}^{\rm T} \right)^{-1} \mathsf{\beta} \mathsf{\gamma}^{-1},
\end{align}
and thus one can write
\begin{equation}
    \mathsf{D} = \mathsf{\Sigma}_{yy}
    + \mathsf{G} \mathsf{\Sigma}_{xx} \mathsf{G}^{\rm T}
    - \mathsf{\Sigma}_{yx}  \mathsf{G}^{\rm T}
    - \mathsf{G} \mathsf{\Sigma}_{xy}.
\end{equation}

\subsection{Profile likelihood}

We now move onto the profile likelihood case. After substituting Eq.~\ref{eq:Taylor series} into Eq.~\ref{eq:general likelihood}, we differentiate with respect to $\bm{\xt}$ to find the maximum likelihood point, $\hat{\bm{x}}_{\rm t}$,
\begin{equation}
    \hat{\bm{x}}_{\rm t} = \hat{\bm{x}}_{\rm o} + \mathsf{C}^{-1} \left( \mathsf{\beta} + \mathsf{G}^{\rm T} \mathsf{\gamma} \right) \left( \bm{\yo} - \bm{f} \left( \bm{\xo} \right) \right).
\end{equation}
One can then substitute this into Eqs.~\ref{eq:general likelihood}\&\ref{eq:Taylor series} to obtain the maximum likelihood value
\begin{equation}
    \label{eq:general profile}
    \begin{split}
        \log \hat{\mathcal{L}} \approx &- \frac{1}{2} \log \det \left( 2 \pi \mathsf{\Sigma}\right) \\
    &- \frac{1}{2} \left(\bm{f} \left( \bm{\xo} \right) - \bm{\yo} \right)^{\rm T} \mathsf{D}^{-1} \left(\bm{f} \left( \bm{\xo} \right) - \bm{\yo} \right).
    \end{split}
\end{equation}
As in the simpler case considered before, we see that the second terms of Eqs.~\ref{eq:general uniform}\&\ref{eq:general profile} are the same, and only the first terms differ.

\subsection{Marginalised Normal Regression}

For the multi-dimensional \mnr{} case, we impose a prior on the $\bm{\xt}$ values to be
\begin{equation}
    P ( \bm{\xt} ) = \frac{1}{\sqrt{\det \left( 2 \pi \mathsf{W} \right)} }
    \exp \left[ - \frac{1}{2} \left( \bm{\xt} - \bm{\mu} \right)^{\rm T} \mathsf{W}^{-1} \left( \bm{\xt} - \bm{\mu} \right) \right],
\end{equation}
where $\mathsf{W}$ is the covariance matrix specifying the distribution of $\bm{\xt}$ parameters; this may be proportional to the identity if the $\bm{\xt}$ are iid draws, however we do not assume this in the following. Using this prior, one must perform the Gaussian integral
\begin{equation}
    \begin{split}
        &P = 
        \frac{1}{\sqrt{\det \left( 2 \pi \mathsf{W} \right)}}  \frac{1}{\sqrt{\det \left( 2 \pi \mathsf{\Sigma} \right)}} \\
        & \times \exp \big[
        - \frac{1}{2} \left( \bm{\mu} - \bm{\xo}, \bm{f} \left( \bm{\xo} \right) + \mathsf{G} \left( \bm{\mu} - \bm{\xo} \right) - \bm{\yo} \right)^{\rm T}
        \mathsf{\Sigma}^{-1} \\
        & \quad 
        \left( \bm{\mu} - \bm{\xo}, \bm{f} \left( \bm{\xo} \right) + \mathsf{G} \left( \bm{\mu} - \bm{\xo} \right) - \bm{\yo} \right)
        \big] \int {\rm d} \bm{\xt} \\
        & \times 
        \exp \left[
         - \frac{1}{2} \left( \bm{\xt} - \bm{\mu}\right)^{\rm T} \mathsf{H} \left( \bm{\xt} - \bm{\mu}\right)
        \right] \\
        & 
        \times \exp \Big[
        - \left[
        \left( \mathsf{\alpha} + \mathsf{G}^{\rm T} \mathsf{\beta}^{\rm T}\right) \left( \bm{\mu} - \bm{\xo} \right) \right. \\
        & + \left. \left( \mathsf{\beta} + \mathsf{G}^{\rm T} \mathsf{\gamma} \right) \left( \bm{f} \left( \bm{\xo} \right) + \mathsf{G} \left( \bm{\mu} - \bm{\xo} \right) - \bm{\yo} \right)
        \right]^{\rm T} \left( \bm{\xt} - \bm{\mu}\right)
        \Big],
    \end{split}
\end{equation}
where we have defined
\begin{equation}
    \mathsf{H} \equiv \mathsf{\alpha} + \mathsf{\beta G} + \mathsf{G}^{\rm T} \mathsf{\beta}^{\rm T} +  \mathsf{G}^{\rm T} \mathsf{\gamma G} + \mathsf{W}^{-1}.
\end{equation}
This yields
\begin{equation}
    \begin{split}
         & \log P = - \frac{1}{2} \log \det \left( 2 \pi \mathsf{M} \right) \\
        & - \frac{1}{2} \left( \bm{\mu} - \bm{\xo}, \bm{f} \left( \bm{\xo} \right) + \mathsf{G} \left( \bm{\mu} - \bm{\xo} \right) - \bm{\yo} \right)^{\rm T} \mathsf{M}^{-1} \\
        & \quad \left( \bm{\mu} - \bm{\xo}, \bm{f} \left( \bm{\xo} \right) + \mathsf{G} \left( \bm{\mu} - \bm{\xo} \right) - \bm{\yo} \right),
    \end{split}
\end{equation}
where
\begin{equation}
    \mathsf{M}^{-1} =
    \begin{pmatrix}
        a & b \\
        b^{\rm T} & c
    \end{pmatrix},
\end{equation}
for
\begin{align}
    a & = \mathsf{\alpha} - \left( \mathsf{\alpha} + \mathsf{\beta G} \right) \mathsf{H}^{-1} \left( \mathsf{\alpha} + \mathsf{G}^{\rm T} \mathsf{\beta}^{\rm T} \right) \\
    b &= \mathsf{\beta} - \left( \mathsf{\alpha} + \mathsf{\beta G} \right) \mathsf{H}^{-1} \left(  \mathsf{\beta} + \mathsf{G}^{\rm T} \mathsf{\gamma} \right) \\
    c &= \mathsf{\gamma} - \left( \mathsf{\beta}^{\rm T} + \mathsf{\gamma G}  \right) \mathsf{H}^{-1} \left( \mathsf{\beta} + \mathsf{G}^{\rm T} \mathsf{\gamma}   \right)
\end{align}
and where the determinant term follows from normalisation requirements. This can be written as
\begin{equation}
    \mathsf{M} = 
    \begin{pmatrix}
        \mathsf{\Sigma}_{xx} + \mathsf{W}&  \mathsf{\Sigma}_{xy} + \mathsf{W} \mathsf{G}^{\rm T} \\
        \mathsf{\Sigma}_{yx} + \mathsf{G W} & \mathsf{\Sigma}_{yy} + \mathsf{G W} \mathsf{G}^{\rm T}
    \end{pmatrix},
\end{equation}
which can be verified by direct substitution. This provides a simple expression for the \mnr{} result with an arbitrary covariance matrix and function being fitted.

Each of these likelihoods can then be sampled or maximised to obtain estimates for the parameters of the function, the intrinsic scatter, and $\mu$ and $w$ in the case of \mnr. The sampling of the parameters of arbitrary functions with arbitrary covariances is implemented in \textsc{roxy}.
In the case of diagonal covariance matrices 
$\mathsf{\Sigma}_{xx} = \diag \left( {\sigma_x^{(1)}}^2,  {\sigma_x^{(2)}}^2, \ldots,  {\sigma_x^{(N)}}^2\right)$, 
$\mathsf{\Sigma}_{yy} = \diag \left( {s_y^{(1)}}^2,  {s_y^{(2)}}^2, \ldots,  {s_y^{(N)}}^2\right)$, $\mathsf{\Sigma}_{xy} = 0$, and where $\yt^{(i)}$ only depends on $\xt^{(i)}$, these results simplify to those in Sections~\ref{sec:unif},\ref{sec:prof}\&\ref{sec:MNR}, but with $A$ and $B$ replaced by
\begin{equation}
    \mathcal{A}_i \equiv  \left. \frac{\partial}{\partial x} f \left( x, \bm{\theta}\right) \right|_{x = x_{\rm o}^{(i)}}, \quad 
    \mathcal{B}_i \equiv f \left( x_{\rm o}^{(i)}, \bm{\theta}\right) - \mathcal{A}_i x_{\rm o}^{(i)}.
\end{equation}

\section{\textsc{roxy}: a public, simple and efficient implementation of MNR}
\label{sec:roxy}

We design and release an implementation of \mnr, dubbed \textsc{roxy} (\emph{Regression and Optimisation with X and Y errors}). This Python code uses automatic differentiation enabled by Jax both to sample the likelihood using Hamiltonian Monte Carlo and to compute the derivative of $f(\bm{x},\bm{\theta})$ (cf. Eq.~\ref{eq:G definition}) required in the likelihood if $f$ is nonlinear.
We employ the NUTS method implemented in \textsc{numpyro},
and assume uniform priors on all parameters in $\bm{\theta}$, although their range can readily be altered by the user.
\textsc{roxy} additionally calculates the maximum likelihood point using Nelder--Mead algorithm \citep{Gao2012}, in case one wants only point estimates of the parameters.
As well as \mnr{} and the GMM method, \textsc{roxy} implements the \unif{} and \prof{} likelihoods, although, in light of the results of Secs.~\ref{sec:setup} and \ref{sec:mock}, these are not recommended in general.

By using the NUTS sampler, \textsc{roxy} is able to produce MCMC chains very quickly. For the galaxy cluster example in Section~\ref{sec:applications} (which contains over 250 data points), a single chain run on a laptop performs approximately 3500 iterations per second, such that a chain with 700 warm-up steps and 5000 samples takes approximately 1.6 seconds to sample. Over 3500 effective samples are produced for each parameter, with Gelman Rubin statistics equal to unity within less than $10^{-2}$. Given its efficiency and simplicity to use (one just needs to define the function to fit, the parameters to sample and their prior ranges), we recommend \textsc{roxy} not just in the presence of $x$ errors (in which case \mnr{} should be used), but also in the absence of $x$ errors. In this case, the \unif{} method with $\sigma_x=0$ is more efficient because \mnr{} would still sample the (unconstrained) extra hyperparameters describing the Gaussian prior.

\textsc{roxy} contains a number of failsafes designed to mitigate the risk of failure during sampling, raising warnings if the number of effective samples for any of the parameters is less that 100 and if the peak of the posterior is near the edge of the prior.
This likely indicates a pathology in the data set or that the prior range is too small.
A warning is also raised if the function's second derivative times the $x$-error size exceeds the first derivative at any point, which indicates that the first-order Taylor expansion (Eq.~\ref{eq:Taylor series}) may be unreliable. As higher order approximations are computationally intractable, we suggest the user to reparametrise the data to make it more linear in this case.

A property of the \mnr{} (as well as \unif{} and \prof{}) likelihood is asymmetry between $x$ and $y$, so that the regression result depends on which variable is considered independent. Ideally the direction of regression would match the direction of causality in the problem. This may be assessed by
treating the scatter of the points around the best-fit line as an \emph{additive noise model}, in which case the independent variable may be identified as the one that has least correlation with the residuals of the fit~\citep{causal_3, causal_4, causal_1, causal_2}. The \textsc{roxy} function \textsc{assess\_causality} fits both the forward and inverse relations to the dataset, produces plots of the data in both directions with both regression models overlaid and the corresponding normalised residuals plotted against the ``$x$'' variable, and calculates various correlation coefficients.
From these it makes a recommendation as to which variable to treat as independent.
We caution however that the correlations may be non-monotonic and/or nonlinear, and therefore recommend picking the regression direction that visually produces the least correlation in the residuals.
Typically the correct regression direction also produces the better-looking fit to the data in both directions, although in some cases none of these diagnostics are conclusive.
Further information and an example of the procedure may be found in the online documentation.

The code is released at \url{https://github.com/DeaglanBartlett/roxy}, with documentation at \url{https://roxy.readthedocs.io}. As well as returning posterior samples and allowing likelihood computations (which can be integrated into the user's larger code), \textsc{roxy} is interfaced with \textsc{arviz} \citep{arviz_2019} to produce trace plots, \textsc{corner} \citep{corner} and \textsc{getdist} \citep{getdist} to make two-dimensional posterior plots, and \textsc{fgivenx} \citep{fgivenx} for posterior predictive plots. Users are welcome to contact the authors for assistance and to propose modifications and additional features.

\section{Discussion and Conclusion}
\label{sec:conc}

This paper analyses the ostensibly simple problem of fitting a function to data with errors on both the dependent and independent variables. A seemingly sensible approach is to use a Gaussian likelihood with a mean given by $f(\xo, \bm{\theta})$ and a variance $\sigma_y^2 + f^\prime(\xo, \bm{\theta})^2 \sigma_x^2$, which corresponds to marginalising over the latent ``true'' $x$ values, $\bm{\xt}$, under the assumption of an infinite uniform prior. We have shown analytically and numerically that this method yields biased results, demonstrating that this choice of prior on $\bm{\xt}$ is inappropriate. We also show that simply maximising the likelihood with respect to the $\bm{\xt}$ values (which is approximately equivalent to marginalising over them with a Jeffreys prior) also gives biased results, which is particularly noticeable when attempting to infer the relation's intrinsic scatter.

We find that the only choice of prior that gives consistently unbiased results is a mixture of one or more Gaussians with means, widths and weights inferred simultaneously with the parameters of the function being fitted. We show that typically one Gaussian is sufficient, and dub the corresponding inference method \emph{Marginalised Normal Regression} (\mnr). We show analytically that for infinite sample sizes with constant uncertainties this gives perfectly unbiased results regardless of the true distribution of $\bm{\xt}$ values, and numerically that this approximately holds also for finite sample sizes and varying uncertainties.

We publicly release a simple and efficient Python implementation of \mnr{} called \textsc{roxy} (Regression and Optimisation with X and Y errors), which is built on Jax \citep{jax2018github} and \textsc{numpyro} \citep{phan2019composable,bingham2019pyro} to allow efficient sampling using the No U-Turns method of Hamiltonian Monte Carlo. This code can handle both linear and  non-linear functions and correlated errors between all the observed $x$ and $y$ values (i.e. arbitrary covariance matrices).

To demonstrate the importance of regression method, we fit 
the scaling between the dynamically estimated masses of cluster and those inferred from the Sunyaev-Zel'dovich effect. 
The
slope of the linear fit is biased by 4.3$\sigma$ and 6.4$\sigma$ for the uniform marginalisation and profile likelihood methods respectively, and the parameter controlling the normalisation by 2.5$\sigma$ and 2.1$\sigma$. The profile likelihood also drastically underestimates the relation's intrinsic scatter.
The MCMC chains used for  
this example
took only a few seconds to run with \textsc{roxy}.

We identify four ways in which this work could be extended. First, although our method works for arbitrary functional forms our bias investigation has been restricted to linear regression; it would be useful to extend this to the fitting of nonlinear functions and assess the impact of $N_g$ in the GMM model in this case.
Second, it would be useful to extend \mnr{} to arbitrary non-Gaussian (e.g. asymmetric) uncertainties, which are sometimes encountered in real-world applications.
Third, as discussed in Sec.~\ref{sec:roxy} our formulation of the problem has an intrinsic asymmetry between the treatment of $x$---which has corresponding latent variables $\xt$---and $y$, which is said to be determined by $x$ up to the uncertainties and intrinsic scatter.
While relations in science are indeed typically assumed to be causal with the independent variable determining the dependent, and hence possess this asymmetry, there may be cases where a more symmetric treatment is desirable. An example would be if the intrinsic scatter is assumed to be Gaussian \emph{perpendicular} to the fitted curve rather than in a fixed (e.g. $y$) direction. A more general method capable of addressing this should be developed.
Finally, additional capabilities that could on occasion be useful are the ability to model outliers as a separate population to inliers, and the ability to handle missing or truncated data.

Given the strong biases incurred when assuming a Jeffreys or infinite uniform prior on $\bm{\xt}$, or the profile likelihood, such assumptions must be avoided in analyses requiring regression. We have identified and rigorously stress-tested a method for overcoming these biases, Marginalised Normal Regression, and made it easy to implement through our public program
\textsc{roxy}.

\section*{Acknowledgements}

We thank David Alonso, Alan Heavens, Pedro Ferreira, Amelia Ford, Will Handley, Sebastian von Hausegger, Florent Leclercq, Anastasia Ponomareva, Richard Stiskalek, Ewoud Wempe and Tariq Yasin for useful inputs and discussion.

The authors contributed equally to this work. DJB is supported by the Simons Collaboration on ``Learning the Universe.''
HD is supported by a Royal Society University Research Fellowship (grant no. 211046).

This project has received funding from the European Research Council (ERC) under the European Union’s Horizon 2020 research and innovation programme (grant agreement No 693024). This work used the DiRAC Complexity and DiRAC@Durham facilities, operated by the University of Leicester IT Services and Institute for Computational Cosmology, which form part of the STFC DiRAC HPC Facility (www.dirac.ac.uk). This equipment is funded by BIS National E-Infrastructure capital grants ST/K000373/1, ST/P002293/1, ST/R002371/1 and ST/S002502/1, STFC DiRAC Operations grant ST/K0003259/1, and Durham University and STFC operations grant ST/R000832/1. DiRAC is part of the National E-Infrastructure.

For the purpose of open access, the authors have applied a Creative Commons Attribution (CC BY) licence to any Author Accepted Manuscript version arising.

Some of the results in this paper have been derived using the
\textsc{fgivenx} \citep{fgivenx},
\textsc{getdist} \citep{getdist},
Jax \citep{jax2018github},
\textsc{numpyro} \citep{phan2019composable,bingham2019pyro},
\textsc{numpy} \citep{Numpy} and
\textsc{scipy} \citep{Scipy}
packages.

\section*{Data availability}

Our implementation of Marginalised Normal Regression and other likelihoods, \textsc{roxy}, is available at \url{https://github.com/DeaglanBartlett/roxy}.
The galaxy cluster data used in Sec.~\ref{sec:applications} is tabulated in \citet{Aguado-Barahona_2022}.
Other data underlying this article may be shared on request to the authors.

\bibliographystyle{mnras}
\bibliography{references}

\appendix

\section{Analytic maximum-likelihood and bias calculations}
\label{app}

To gain some analytic insight into the results of the various analysis choices, in this section we derive the maximum likelihood values for the parameters of a straight line with intrinsic scatter for each of the methods considered. In this section we assume that $\sigma_x^{(i)}$ and $\sigma_y^{(i)}$ are the same for all $i$, and are equal to $\sigma_x$ and $\sigma_y$, respectively. For convenience, we define $s^2 \equiv \sigma_y^2 + \sint^2$.

We use the following notation
\begin{equation}
    \langle f (\bm{\xo}, \bm{\yo}) \rangle
    \equiv
    \frac{1}{N} \sum_{i=1}^N   f \left(\xo^{(i)}, \yo^{(i)} \right),
\end{equation}
\begin{equation}
    \var \left( \bm{\xo} \right) \equiv \langle \bm{\xo}^2 \rangle - \langle \bm{\xo} \rangle^2,
\end{equation}
and 
\begin{equation}
    \cov \left( \bm{\xo}, \bm{\yo} \right) \equiv \langle \bm{\xo} \bm{\yo} \rangle - \langle \bm{\xo} \rangle \langle \bm{\yo} \rangle.
\end{equation}

For each case considered, we wish to find the bias on the maximum likelihood point. For parameter $Z \in \{A, B, s\}$, this is defined as
\begin{equation}
    \delta \hat{Z} \equiv \lim_{N\to\infty}\hat{Z} - \tilde{Z},
\end{equation}
where, as in the main text, hat denotes maximum-likelihood value, tilde denotes true (generating) value and $N$ is the number of datapoints. The bias values we derive are therefore asymptotic in the limit of much data.

To compute this, it is worth pre-computing a few quantities. We note that we have
\begin{equation}
    \yo^{(i)} = \yt^{(i)} + n_i, \quad
    \yo^{(i)} = \yt^{(i)} + m_i,
\end{equation}
where $n_i$ and $m_i$ are noise terms, such that
\begin{equation}
    \lim_{N \to \infty} \langle n \rangle = \lim_{N \to \infty} \langle m \rangle = 0,
\end{equation}
and
\begin{equation}
    \lim_{N \to \infty} \langle n^2 \rangle = \tilde{s}^2,
    \quad
    \lim_{N \to \infty} \langle m^2 \rangle = \sigma_x^2,
    \quad
    \lim_{N \to \infty} \langle n m \rangle = 0.
\end{equation}
We therefore find
\begin{equation}
    \label{eq:Limit cov xy}
    \lim_{N \to \infty} \cov \left(\bm{\xo}, \bm{\yo} \right)
    = \tilde{A} \var \left( \bm{\xt} \right).
\end{equation}
Similarly,
\begin{equation}
    \label{eq:Limit var x}
    \lim_{N \to \infty} \var \left( \bm{\xo} \right)
    = \var \left( \bm{\xt} \right) + \sigma_x^2,
\end{equation}
and
\begin{equation}
    \label{eq:Limit var y}
    \lim_{N \to \infty} \var \left( \bm{\yo} \right) = 
     \var \left( \bm{\yt} \right) + \tilde{s}^2
     = \tilde{A}^2 \var \left( \bm{\xt} \right) + \tilde{s}^2 .
\end{equation}

We now specialise in turn to the \unif, \prof{} and \mnr{} methods.

\subsection{Infinite uniform marginalisation}

\subsubsection{Maximum likelihood estimates}
\label{sec:app uniform}

If we assume a uniform prior over the $\bm{\xt}$ values, as well as the other parameters of the model, we can marginalise over $\bm{\xt}$ to obtain
\begin{equation}
    \begin{split}
         \log P & \left( A, B, s \vert \bm{\xo}, \bm{\yo} \right)
        = - \frac{N}{2} \log \left(A^2 \sigma_x^2 + s^2 \right) \\
        & - \frac{1}{2} \sum_i \frac{\left( \yo^{(i)} - A \xo^{(i)} - B \right)^2}{A^2 \sigma_x^2 + s^2}
    + {\rm const}.
    \end{split}
\end{equation}
Taking the derivative with respect to $B$ we obtain the maximum likelihood estimate
\begin{equation}
    \label{eq:marg: B deriv}
    \hat{B} = \frac{1}{N} \sum_i \left( \yo^{(i)} - \hat{A} \xo^{i} \right)
    = \langle \bm{\yo} \rangle - \hat{A} \langle \bm{\xo} \rangle.
\end{equation}
Similarly, for $s$ we have
\begin{equation}
    - N \left(\hat{A}^2 \sigma_x^2 + \hat{s}^2 \right) + \sum_i \left( \yo^{(i)} - \hat{A} \xo^{(i)} - \hat{B} \right)^2 = 0,
\end{equation}
and for $A$
\begin{equation}
    \begin{split}
        &\hat{A}\sigma_x^2 \left[ - N \left(\hat{A}^2 \sigma_x^2 + \hat{s}^2 \right) + \sum_i \left( \yo^{(i)} - \hat{A} \xo^{(i)} - \hat{B} \right)^2 \right] \\
        & + \left(\hat{A}^2 \sigma_x^2 + \hat{s}^2 \right) \sum_i \xo^{(i)} \left( \yo^{(i)} - \hat{A} \xo^{(i)} - \hat{B} \right) = 0.
    \end{split}
\end{equation}
By combining these two results, we see that
\begin{equation}
    \hat{B} \sum_i \xo^{(i)} = \sum_i \xo^{(i)} \left( \yo^{(i)} - \hat{A} \xo^{(i)} \right).
\end{equation}
We can then substitute Eq.~\ref{eq:marg: B deriv} to find
\begin{equation}
    \hat{A} = \frac{\cov \left(\bm{\xo}, \bm{\yo} \right)}{\var \left( \bm{\xo} \right)}.
\end{equation}
Substituting into the expression for $\hat{B}$, we then immediately find that
\begin{equation}
    \hat{B} = \frac{\langle \bm{\xo}^2 \rangle \langle \bm{\yo} \rangle - \langle \bm{\xo} \bm{\yo} \rangle \langle \bm{\xo} \rangle}{\var \left( \bm{\xo} \right)}.
\end{equation}
Finally, we can use these two results to obtain the maximum likelihood estimate for the intrinsic scatter. With a little algebraic manipulation, one finds
\begin{equation}
    \hat{s}^2 = 
    \var \left( \bm{\yo} \right)
    - \frac{\cov \left(\bm{\xo}, \bm{\yo} \right)^2}{\var \left( \bm{\xo} \right)^2} \left( \var \left( \bm{\xo} \right) + \sigma_x^2 \right).
\end{equation}
This final expression states that the variance in the $y$ direction is attributed to the variance in $x$ and the errors in $x$ (weighted by the slope of the line), with any additional variance assigned to $\hat{s}$.

\subsubsection{Biases}

We now compute the biases for the three maximum likelihood values we have just computed, using the results in Eqs.~\ref{eq:Limit cov xy},\ref{eq:Limit var x}\&\ref{eq:Limit var y}. For the slope this is
\begin{equation}
    \delta \hat{A} = \frac{\tilde{A} \var \left( \bm{\xt} \right)}{\var \left( \bm{\xt} \right) + \sigma_x^2} - \tilde{A}
    = - \frac{\tilde{A} \sigma_x^2}{\var \left( \bm{\xt} \right) + \sigma_x^2},
\end{equation}
yielding a negative bias (underestimation) if $\tilde{A} > 0$. For the intercept
\begin{equation}
    \delta \hat{B} 
    = \langle \bm{\yt} \rangle - \frac{\tilde{A} \var \left( \bm{\xt} \right)}{\var \left( \bm{\xt} \right) + \sigma_x^2} \langle \bm{\xt} \rangle - \tilde{B} 
    = \frac{\tilde{A} \sigma_x^2 \langle \bm{\xt} \rangle}{\var \left( \bm{\xt} \right) + \sigma_x^2},
\end{equation}
so this is also biased with a sign depending on $\tilde{A} \langle \bm{\xt} \rangle$. Finally, the intrinsic scatter is overestimated:
\begin{equation}
    \begin{split}
        \delta \hat{s}^2 &=
        \tilde{A}^2 \var \left( \bm{\xt} \right) - \frac{\tilde{A}^2 \var \left( \bm{\xt} \right)^2}{\left(\var \left( \bm{\xt} \right) + \sigma_x^2\right)^2} \left(\var \left( \bm{\xt} \right) + 2 \sigma_x^2\right) \\
        &= \frac{\tilde{A}^2 \var \left( \bm{\xt} \right) \sigma_x^4}{\left(\var \left( \bm{\xt} \right) + \sigma_x^2\right)^2}.
    \end{split}
\end{equation}

\subsection{Profile likelihood}
\label{sec:app profile}

\subsubsection{Maximum likelihood estimates}

The profile likelihood begins with the original full likelihood that includes the $\bm{\xt}$:
\begin{equation}
    \label{eq:prof like}
    \begin{split}
        \log P & \left( \bm{\xt}, A, B, s \right)
        =
        - N \log s
        - \sum_i \frac{\left( \xo^{(i)} - \xt^{(i)}\right)^2}{2 \sigma_x^2} \\
        & - \sum_i \frac{\left( \yo^{(i)} - A \xt^{(i)} - B \right)^2}{2 s^2} + {\rm const}.
    \end{split}
\end{equation}
We begin by optimising this with respect to $\xt^{(i)}$ to obtain
\begin{equation}
    \label{eq:prof xt deriv}
    \hat{x}_{\rm t}^{(i)}
    =
    \frac{s^2}{s^2 + A^2 \sigma_x^2} \xo^{(i)} 
    + \frac{A\sigma_x^2}{s^2 + A^2 \sigma_x^2} \left( \yo^{(i)} - B \right),
\end{equation}
which is the maximum likelihood $\hat{x}_{\rm t}^{(i)}$ given $A$, $B$ and $s$.
Substituting this back into Eq.~\ref{eq:prof like} and removing the arbitrary constant, we have
\begin{equation}
    \label{eq:prof L3}
    \log \mathcal{L}_3 (A, B, s) = 
    - N \log s - \frac{1}{2} \sum_i \frac{\left( \yo^{(i)} - A \xo^{(i)} - B \right)^2}{s^2 + A^2 \sigma_x^2},
\end{equation}
where $\mathcal{L}_3$ is $P$ evaluated at the maximum likelihood $\bm{\xt}$, and the subscript 3 denotes that it has three arguments. We can then optimise with respect to $B$ to obtain
\begin{equation}
    \label{eq:prof B result}
    \hat{B} = \langle \bm{\yo} \rangle - A \langle \bm{\xo} \rangle.
\end{equation}
Substituting this into Eq.~\ref{eq:prof L3}, we have
\begin{equation}
    \label{eq:prof L2}
    \begin{split}
        \log \mathcal{L}_2 (A, s) &=
        - N \log s \\
        & - \frac{N}{2} \frac{\var \left( \bm{\yo} \right) + A^2 \var \left( \bm{\xo} \right) - 2 A \cov \left( \bm{\xo}, \bm{\yo} \right)}{s^2 + A^2 \sigma_x^2}.
    \end{split}
\end{equation}

We now need to optimise Eq.~\ref{eq:prof L2} with respect to $A$ and $s$ to find $\hat{A}$ and $\hat{s}$. 
Since $s^2 = \sigma_y^2 + \sint^2$ where both $\sigma_y$ and $\sint$ are real, we need $\hat{s}^2 \geq \sigma_y^2$ for a valid solution. The global maximum of the likelihood does not necessarily obey this condition, and, in that case, the parameters which maximise the likelihood within the allowed range of $\hat{s}$ may not be a turning point of the likelihood. So, although we now differentiate with respect to $s$ to find the global maximum, it is important to check whether this $s$ is permitted and also compare to the likelihood at $s=\sigma_y$.
Optimising with respect to $s$ gives
\begin{equation}
    \label{eq:prof s deriv}
    \begin{split}
        &\left( \hat{s}^2 + A^2 \sigma_x^2 \right)^2 \\
        &= s^2
        \left( \var \left( \bm{\yo} \right) + A^2 \var \left( \bm{\xo} \right) - 2 A \cov \left( \bm{\xo}, \bm{\yo} \right) \right),
    \end{split}
\end{equation}
and with respect to $A$
\begin{equation}
    \label{eq:prof A deriv}
    \begin{split}
        &\left( \hat{A} \var \left( \bm{\xo} \right) - \cov \left( \bm{\xo}, \bm{\yo} \right) \right) \left( s^2 + \hat{A}^2 \sigma_x^2 \right) \\
        &= 
        \hat{A} \sigma_x^2 \left( \var \left( \bm{\yo} \right) + \hat{A}^2 \var \left( \bm{\xo} \right) - 2 \hat{A} \cov \left( \bm{\xo}, \bm{\yo} \right) \right).
    \end{split}
\end{equation}
Eqs.~\ref{eq:prof A deriv}\&\ref{eq:prof s deriv} have the same factor on the right hand side, so we can combine them to obtain
\begin{equation}
    \hat{A} \sigma_x^2 \left( \hat{s}^2 + \hat{A}^2 \sigma_x^2 \right) =
    \hat{s}^2  \left( \hat{A} \var \left( \bm{\xo} \right) - \cov \left( \bm{\xo}, \bm{\yo} \right) \right).
\end{equation}
This can then be combined with Eq.~\ref{eq:prof A deriv} to get (if $\hat{A} \neq 0$)
\begin{equation}
    \label{eq:prof A s intermediate}
    \hat{A}^2 \sigma_x^2 + \hat{s}^2 = \var \left( \bm{\yo} \right) - \hat{A} \cov \left( \bm{\xo}, \bm{\yo} \right).
\end{equation}
Repeated substitution of Eq.~\ref{eq:prof A s intermediate} into Eq.~\ref{eq:prof s deriv} yields (if $\hat{A} \neq 0$)
\begin{equation}
    \hat{A} = \frac{\cov \left( \bm{\xo}, \bm{\yo} \right) \left( \var \left( \bm{\yo} \right) - 2 \hat{s}^2 \right)}{\cov \left( \bm{\xo}, \bm{\yo} \right)^2 - \hat{s}^2 \var \left( \bm{\xo} \right) + \sigma_x^2 \var \left( \bm{\yo} \right)}.
\end{equation}
We now substitute this into Eq.~\ref{eq:prof A s intermediate} to find that $u \equiv \hat{s}^2$ must solve the cubic
\begin{equation}
    \label{eq:prof cubic}
    a u^3 + b u^2 + c u + d = 0,
\end{equation}
where
\begin{equation}
    a = \var \left( \bm{\xo} \right)^2,
\end{equation}
\begin{equation}
    \begin{split}
         b =& - \var \left( \bm{\xo} \right)^2 \var \left( \bm{\yo} \right) + 4 \cov \left( \bm{\xo}, \bm{\yo} \right)^2 \sigma_x^2 \\
         &- 2 \var \left( \bm{\xo} \right) \var \left( \bm{\yo} \right) \sigma_x^2,
    \end{split}
\end{equation}
\begin{equation}
    \begin{split}
    c &= - \cov \left( \bm{\xo}, \bm{\yo} \right)^4 + \cov \left( \bm{\xo}, \bm{\yo} \right)^2 \var \left( \bm{\xo} \right) \var \left( \bm{\yo} \right) \\
    &- 4 \cov \left( \bm{\xo}, \bm{\yo} \right)^2 \var \left( \bm{\yo} \right)  \sigma_x^2 \\
    &\quad + 2  \var \left( \bm{\xo} \right)  \var \left( \bm{\yo} \right)^2 \sigma_x^2 + \var \left( \bm{\yo} \right)^2 \sigma_x^4, \end{split}
\end{equation}
\begin{equation}
    d = - \var \left( \bm{\yo} \right)^3 \sigma_x^4.
\end{equation}

The solutions for this cubic are outlined in Algorithm~\ref{alg:cubic real roots}, and derived in Section~\ref{sec:cubic properties}. The parameters of interest for us are
\begin{equation}
    \label{eq:D0 prof}
    \begin{split}
    &\Delta_0 =
    \var \left( \bm{\xo} \right)^2  \var \left( \bm{\yo} \right)^2 (\var \left( \bm{\xo} \right) - \sigma_x^2)^2 \\
    & + \cov \left( \bm{\xo}, \bm{\yo} \right)^2 \var \left( \bm{\xo} \right)  \var \left( \bm{\yo} \right) \\
    & \times (-3 \var \left( \bm{\xo} \right)^2 + 4 \var \left( \bm{\xo} \right) \sigma_x^2 - 16 \sigma_x^4) \\
    & + \cov \left( \bm{\xo}, \bm{\yo} \right)^4 (3 \var \left( \bm{\xo} \right)^2 + 16 \sigma_x^4),
    \end{split}
\end{equation}
\begin{equation}
    \label{eq:D1 prof}
    \begin{split}
        &\Delta_1 =
        -2 \var \left( \bm{\xo} \right)^3 \var \left( \bm{\yo} \right)^3 (\var \left( \bm{\xo} \right) - \sigma_x^2)^3 \\
        & + 3 \cov \left( \bm{\xo}, \bm{\yo} \right)^2 \var \left( \bm{\xo} \right)^2 \var \left( \bm{\yo} \right)^2 \\
        & \times (3 \var \left( \bm{\xo} \right)^3 + 2 \var \left( \bm{\xo} \right)^2 \sigma_x^2 - 16 \var \left( \bm{\xo} \right) \sigma_x^4 + 20 \sigma_x^6) \\
        & - 3 \cov \left( \bm{\xo}, \bm{\yo} \right)^4 \var \left( \bm{\xo} \right) \var \left( \bm{\yo} \right) \\
        & \times (3 \var \left( \bm{\xo} \right)^3 + 18 \var \left( \bm{\xo} \right)^2 \sigma_x^2 - 16 \var \left( \bm{\xo} \right) \sigma_x^4 + 64 \sigma_x^6) \\
        & + 4 \cov \left( \bm{\xo}, \bm{\yo} \right)^6 (9 \var \left( \bm{\xo} \right)^2 \sigma_x^2 + 32 \sigma_x^6), 
    \end{split}
\end{equation}
defined in Eq.~\ref{eq:cubic param definitions}.
One cannot significantly simplify these expressions to obtain a simple, compact form for $\hat{s}^2$. Moreover, depending on the data set studied, one may need to use the different solutions given in Algorithm~\ref{alg:cubic real roots}, and thus there is not a single set of solutions which applies in all cases.

\subsubsection{Biases}

Given that there does not exist a single, simple expression for the inferred values of $\hat{A}$, $\hat{B}$ and $\hat{s}^2$, we cannot compute general closed-form expressions for their biases. 
Since many of the terms used to compute the various solutions for $\hat{s}^2$ when using Eqs.~\ref{eq:D0 prof}\&\ref{eq:D1 prof} do not cancel (see Algorithm~\ref{alg:cubic real roots} and Section \ref{sec:cubic properties}), we find $\hat{s}\neq\tilde{s}$ and hence that this method is biased.
For a quantification of this, it is more instructive to consider the more general numerical experiments of Section~\ref{sec:mock}, rather than these analytic expressions.

\subsection{Gaussian hyperprior}
\label{sec:app mnr}

\subsubsection{Maximum likelihood estimates}

We now assume the $\bm{\xt}$ are drawn from a Gaussian distribution of mean $\mu$ and width $w$, where these hyperparameters themselves have uniform priors. In this case, the marginalisation over $\bm{\xt}$ gives
\begin{equation}
    \begin{split}
        & \log P \left( A, B, s, \mu, w \vert \bm{\xo}, \bm{\yo} \right) \\
    & = - \frac{1}{2} \sum_i \frac{ w^2 \left( A \xo^{(i)} + B - \yo^{(i)} \right)^2 + \sigma_x^2 \left( A \mu + B - \yo^{(i)} \right)^2}{A^2 w^2 \sigma_x^2 + s^2 \left( w^2 + \sigma_x^2 \right)} \\
    & - \frac{1}{2} \sum_i \frac{s^2 \left( \xo^{(i)} - \mu \right)^2}{A^2 w^2 \sigma_x^2 + s^2 \left( w^2 + \sigma_x^2 \right)} \\
    & - \frac{N}{2} \log \left(A^2 w^2 \sigma_x^2 + s^2 \left( w^2 + \sigma_x^2 \right) \right)+ {\rm const}.
    \end{split}
\end{equation}

Let us begin by maximising this likelihood with respect to $\mu$ to give
\begin{equation}
    \hat{\mu} \sum_i \left(\hat{A}^2 \sigma_x^2 + \hat{s}^2 \right)
    = \hat{s}^2 \sum_i \xo^{(i)} + \hat{A}\sigma_x^2 \sum_i \left( \yo^{(i)} - \hat{B} \right),
\end{equation}
and with respect to $B$ to give
\begin{equation}
    \hat{B} = \langle \bm{\yo} \rangle - \frac{\hat{A}}{\hat{w}^2 + \sigma_x^2} \left( \hat{w}^2 \langle \bm{\xo} \rangle + \sigma_x^2 \hat{\mu} \right).
\end{equation}
By substituting this value of $\hat{B}$ into the equation for $\hat{\mu}$, we find that many terms cancel such that
\begin{equation}
    \label{eq:mnr mu result}
    \hat{\mu} = \langle \bm{\xo} \rangle,
\end{equation}
independent of all other parameters. Substituting this back into our expression for $\hat{B}$, we have
\begin{equation}
    \label{eq:mnr B result}
    \hat{B} = \langle \bm{\yo} \rangle - \hat{A} \langle \bm{\xo} \rangle.
\end{equation}

To make further progress, we must now maximise with respect to $A$, $w$ and $s$. Doing this, we find for $A$
\begin{equation}
    \label{eq:mnr A deriv}
    \begin{split}
        \hat{A} & \hat{w}^2 \sigma_x^2 K
        =
        \left( \hat{A}^2 \hat{w}^2 \sigma_x^2 + \hat{s}^2 \left( \hat{w}^2 + \sigma_x^2 \right) \right) \\
        & \times \sum_i \left[
        \hat{w}^2 \left( \hat{A} \xo^{(i)} + \hat{B} - \yo^{(i)} \right) \xo^{(i)} \right. \\
        & \left. \qquad \qquad + \sigma_x^2 \left( \hat{A} \hat{\mu} + \hat{B} - \yo^{(i)} \right) \hat{\mu}
    \right],
    \end{split}
\end{equation}
then for $s$
\begin{equation}
    \label{eq:mnr s deriv}
    \left( \hat{w}^2 + \sigma_x^2\right)K
    =
    \left( \hat{A}^2 \hat{w}^2 \sigma_x^2 + \hat{s}^2 \left( \hat{w}^2 + \sigma_x^2 \right) \right)
    \sum_i 
    \left( \xo^{(i)} - \hat{\mu}\right)^2,
\end{equation}
and finally for $w$
\begin{equation}
    \label{eq:mnr w deriv}
    \begin{split}
         &\left( \hat{A}^2 \sigma_x^2 + \hat{s}^2 \right) K =  \\
        & \left( \hat{A}^2 \hat{w}^2 \sigma_x^2 + \hat{s}^2 \left( \hat{w}^2 + \sigma_x^2 \right) \right)
        \sum_i 
        \left( \hat{A} \xo^{(i)} + \hat{B} - \yo^{(i)} \right)^2,
    \end{split}
\end{equation}
where
\begin{equation}
    \begin{split}
        K \equiv& - N \left(\hat{A}^2 \hat{w}^2 \sigma_x^2 + \hat{s}^2 \left( \hat{w}^2 +\sigma_x^2 \right) \right) \\
        &+
        \sum_i \hat{w}^2 \left( \hat{A} \xo^{(i)} + \hat{B} - \yo^{(i)} \right)^2 \\
        &+ \sum_i\sigma_x^2 \left( \hat{A}\hat{\mu} + \hat{B} - \yo^{(i)} \right)^2 
        + \sum_i \hat{s}^2 \left( \xo^{(i)} - \hat{\mu} \right)^2.
    \end{split}
\end{equation}
Taking the ratio of Eqs.~\ref{eq:mnr A deriv}\&\ref{eq:mnr s deriv} and using Eqs.~\ref{eq:mnr mu result}\&\ref{eq:mnr B result}, we find
\begin{equation}
    \label{eq:mnr A intermediate}
    \hat{A} = 
    \frac{\hat{w}^2 + \sigma_x^2}{\hat{w}^2}
    \frac{\cov \left(\bm{\xo}, \bm{\yo} \right)}{\var \left( \bm{\xo} \right)}.
\end{equation}
Similarly, taking the ratio of Eqs.~\ref{eq:mnr s deriv}\&\ref{eq:mnr w deriv} and using Eqs.~\ref{eq:mnr mu result}\&\ref{eq:mnr B result}, we have
\begin{equation}
    \label{eq:mnr s intermediate}
    \begin{split}
        \hat{s}^2 = 
        \frac{\hat{w}^2 + \sigma_x^2 }{\hat{w}^2 \var \left( \bm{\xo} \right)^2} &
        \left[ 
        \hat{w}^2 \var \left( \bm{\xo} \right) \var \left( \bm{\yo} \right) \right. \\
        & \left. - \left(\hat{w}^2 + \sigma_x^2 \right)^2 \cov \left( \bm{\xo}, \bm{\yo} \right)^2
    \right].
    \end{split}
\end{equation}
Before using these results to obtain $\hat{w}$, we rewrite Eq.~\ref{eq:mnr s deriv} as
\begin{equation}
    \begin{split}
        &\left( \hat{w}^2 + \sigma_x^2\right)^2 \left(\var \left( \bm{\yo} \right) - \hat{s}^2 \right) 
        - 2 \hat{A} \hat{w}^2 \left( \hat{w}^2 + \sigma_x^2\right) \cov \left( \bm{\xo}, \bm{\yo} \right) \\
        & + \hat{A}^2 \hat{w}^2 \left[ \hat{w}^2 \var \left( \bm{\xo} \right) - \left( \hat{w}^2 + \sigma_x^2\right) \sigma_x^2 \right]  = 0.
    \end{split}
\end{equation}
Then, upon substitution of Eqs.~\ref{eq:mnr A intermediate}\&\ref{eq:mnr s intermediate} we obtain an expression for $\hat{w}$
\begin{equation}
    \label{eq:mnr w result}
    \hat{w}^2 =
    \var \left( \bm{\xo} \right) - \sigma_x^2.
\end{equation}
This can be substituted back into Eqs.~\ref{eq:mnr A intermediate}\&\ref{eq:mnr s intermediate} to find
\begin{equation}
    \label{eq:mnr A result}
    \hat{A} = \frac{\cov \left( \bm{\xo}, \bm{\yo} \right)}{\var \left( \bm{\xo} \right) - \sigma_x^2},
\end{equation}
and
\begin{equation}
    \label{eq:mnr s result}
    \hat{s}^2 
    =
    \var \left( \bm{\yo} \right)
    -
    \frac{\cov \left( \bm{\xo}, \bm{\yo} \right)^2}{\var \left( \bm{\xo} \right) - \sigma_x^2}
\end{equation}
One can then substitute Eq.~\ref{eq:mnr A result} into Eq.~\ref{eq:mnr B result} to obtain $\hat{B}$.

\subsubsection{Biases}

We now compute the biases for these three maximum likelihood values, using the results in Eqs.~\ref{eq:Limit cov xy},\ref{eq:Limit var x}\&\ref{eq:Limit var y}. The bias in the slope is
\begin{equation}
    \delta \hat{A} = 
    \frac{\tilde{A} \var \left( \bm{\xt} \right)}{\var \left( \bm{\xt} \right) + \sigma_x^2 - \sigma_x^2} - \tilde{A} = 0,
\end{equation}
so $\hat{A}$ gives the truth as $N \to \infty$. Now the intercept is (using that $\hat{A} = \tilde{A}$ in this limit)
\begin{equation}
    \delta \hat{B} = \langle \bm{\yt} \rangle - \tilde{A} \langle \bm{\xt} \rangle - \tilde{B} = 0,
\end{equation}
so this also gives the correct result. Finally, for the variance we get
\begin{equation}
    \delta \hat{s}^2 = 
    \tilde{A}^2 \var \left( \bm{\xt} \right) + \tilde{s}^2 - \frac{\tilde{A}^2 \var \left( \bm{\xt} \right)^2}{\var \left( \bm{\xt} \right) + \sigma_x^2 - \sigma_x^2} - \tilde{s}^2 = 0.
\end{equation}
We therefore see that the maximum-likelihood values of all three linear regression parameters equal their true values in \mnr. This method is therefore the only one that is asymptotically fully unbiased.

\section{Properties of cubic equations}
\label{sec:cubic properties}

When using the profile likelihood in Section~\ref{sec:app profile}, we had to find the roots of a cubic equation (Eq.~\ref{eq:prof cubic}). To know whether or not we have a bias for this method, we must find the real roots of this equation (since our parameters are real) so that we can compute the method's asymptotic bias. It is therefore instructive to investigate some properties of the roots of such equations, which may prove useful for other investigations also. In this section we will solve the equation
\begin{equation}
    \label{eq:General cubic}
    a u^3 + b u^2 + c u + d = 0,
\end{equation}
for $u$, where $a, b, c, d \in \mathbb{R}$, defining for convenience
\begin{equation}
    \label{eq:cubic param definitions}
    \Delta_0  \equiv b^2 - 3ac, \quad
    \Delta_1  \equiv 2 b^3 - 9abc + 27 a^2 d, \quad
    \xi \equiv e^{2 \pi \iu / 3}.
\end{equation}
\subsection{First set of solutions}

We define
\begin{equation}
    C_+ \equiv \left( \frac{\Delta_1 + \sqrt{\Delta_1^2 - 4 \Delta_0^3}}{2} \right)^{1/3},
\end{equation}
and assume in this subsection than $C_+ \neq 0$. It will be convenient to further define
\begin{equation}
    Q_+ \equiv \frac{\Delta_1 + \sqrt{\Delta_1^2 - 4 \Delta_0^3}}{2} = C_+^3.
\end{equation}
Then the three solutions of the cubic are
\begin{equation}
    \label{eq:Cubic 1 solutions}
    u_k = - \frac{1}{3a} \left( b + \xi^k C_+ + \frac{\Delta_0}{\xi^k C_+} \right), \quad k \in \{0, 1, 2 \}.
\end{equation}
If we define $C_+ \equiv r_+ e^{\iu \theta_+}$ for $r_+, \theta_+ \in \mathbb{R}$ and $r_+ > 0$, then this becomes
\begin{equation}
    u_k = - \frac{1}{3a} \left( b + \left(r_+ + \frac{\Delta_0}{r_+} \right) \cos \phi_k +  \iu \left(r_+ - \frac{\Delta_0}{r_+} \right) \sin \phi_k \right),
\end{equation}
where
\begin{equation}
    \phi_0 = \theta_+, \quad
    \phi_1 = \theta_+ + \frac{2 \pi}{3}, \quad
    \phi_2 = \theta_+ - \frac{2 \pi}{3}.
\end{equation}
For $u_k$ to be real (as we will require), we have two options:
\begin{equation}
    \label{eq:Cubic 1 real solution conditions}
    r_+^2 = \Delta_0 \quad
    \text{or} \quad
    \sin \phi_k = 0.
\end{equation}

\subsubsection{Case 1}

For the first case, let us assume that $\Delta_1^2 \leq 4 \Delta_0^3$. In this case, 
\begin{equation}
    |Q_+|^2  = \frac{\Delta_1^2}{4} + \frac{4 \Delta_0^3 - \Delta_1^2}{4} = \Delta_0^3 \implies
    r_+^2 = |C_+|^2 = \Delta_0.
\end{equation}
Thus, the first condition of Eq.~\ref{eq:Cubic 1 real solution conditions} is satisfied, and thus $u_0$, $u_1$ and $u_2$ are all real if $\Delta_1^2 \leq 4 \Delta_0^3$.

\subsubsection{Case 2}

Here we consider the case $\Delta_1^2 > 4 \Delta_0^3$, so that $C_+ \in \mathbb{R}$ and $\theta_+ = 0$. Hence, $\phi_0 = 0$ and thus $u_0$ is always real (according to Eq.~\ref{eq:Cubic 1 real solution conditions}).

For $k=1,2$ we see that we require $r^2 = \Delta_0$. Since $C_+ \in \mathbb{R}$, we have $r_+^2 = C_+^2$ and thus we require
\begin{equation}
    \begin{split}
        & \left( \Delta_1 + \sqrt{\Delta_1^2 - 4 \Delta_0^3} \right)^2 = 4 \Delta_0^3 \\
        & \implies
        \sqrt{\Delta_1^2 - 4 \Delta_0^3} \left( \Delta_1 + \sqrt{\Delta_1^2 - 4 \Delta_0^3} \right) = 0.
    \end{split}
\end{equation}
By our assumption for this branch of solutions, the first term cannot be zero. Thus, we require the term in the brackets to be zero for $u_1$ and/or $u_2$ to be real. However, this would make $C_+=0$ which is not a valid for this set of solutions. Hence, $u_0$ and $u_1$ cannot be real for $\Delta_1^2 > 4 \Delta_0^3$.

We therefore conclude for this set of solutions, that $u_0$ is always real if $\Delta_1^2 > 4 \Delta_0^3$, whereas $u_1$ and $u_2$ can never be real under this assumption.

\subsection{Second set of solutions}

If $C_+=0$, then we must consider a second set of solutions. We define
\begin{equation}
    C_- \equiv \left( \frac{\Delta_1 - \sqrt{\Delta_1^2 - 4 \Delta_0^3}}{2} \right)^{1/3},
\end{equation}
and assume in this subsection than $C_- \neq 0$. Similarly to before, we define
\begin{equation}
    Q_- \equiv \frac{\Delta_1 - \sqrt{\Delta_1^2 - 4 \Delta_0^3}}{2} = C_-^3.
\end{equation}
Then the three solutions of the cubic are
\begin{equation}
    \label{eq:Cubic 2 solutions}
    v_k = - \frac{1}{3a} \left( b + \xi^k C_- + \frac{\Delta_0}{\xi^k C_-} \right), \quad k \in \{0, 1, 2 \}.
\end{equation}
By the same reasoning as above, we define $C_- \equiv r_- e^{\iu \theta_+}$ for $r_-, \theta_- \in \mathbb{R}$ and $r_- > 0$, and thus find the condition for $v_k$ to be real is
\begin{equation}
    \label{eq:Cubic 2 real solution conditions}
    r_-^2 = \Delta_0 \quad
    \text{or} \quad
    \sin \varphi_k = 0,
\end{equation}
where
\begin{equation}
    \varphi_0 = \theta_-, \quad
    \varphi_1 = \theta_- + \frac{2 \pi}{3}, \quad
    \varphi_2 = \theta_- - \frac{2 \pi}{3}.
\end{equation}

\subsubsection{Case 1}

We again begin with the case $\Delta_1^2 \leq 4 \Delta_0^3$. In this case, 
\begin{equation}
    |Q_-|^2  = \frac{\Delta_1^2}{4} + \frac{4 \Delta_0^3 - \Delta_1^2}{4} = \Delta_0^3 \implies
    r_-^2 = |C_-|^2 = \Delta_0.
\end{equation}
Thus, the first condition of Eq.~\ref{eq:Cubic 2 real solution conditions} is satisfied, and thus $v_0$, $v_1$ and $v_2$ are all real if $\Delta_1^2 \leq 4 \Delta_0^3$.

\subsubsection{Case 2}

Now moving onto the second case where $\Delta_1^2 > 4 \Delta_0^3$, we again note that this makes $C_- \in \mathbb{R}$ and thus $\theta_- =0$. Hence, $\varphi_0=0$ and so, by the second condition in Eq.~\ref{eq:Cubic 2 real solution conditions}, we find that $v_0$ is real.

For $v_1$ and $v_2$, we again must consider the first condition of Eq.~\ref{eq:Cubic 2 real solution conditions}. This requires $C_-^2 = r_-^2 = \Delta_0$, and hence we need
\begin{equation}
    \begin{split}
        &\left( \Delta_1 - \sqrt{\Delta_1^2 - 4 \Delta_0^3} \right)^2 = 4 \Delta_0^3 \\
        & \implies
        \sqrt{\Delta_1^2 - 4 \Delta_0^3} \left( \Delta_1 - \sqrt{\Delta_1^2 - 4 \Delta_0^3} \right) = 0.
    \end{split}
\end{equation}
Once again, we find that the first term cannot be zero based on our assumptions. This requires the bracketed term to be zero, but this makes $C_-=0$, which again is not valid for this set of solutions. 

Hence, under the assumption $\Delta_1^2 > 4 \Delta_0^3$, we find that $v_0$ is always real, whereas $v_1$ and $v_2$ are never real.

\subsection{Third set of solutions}

If $\Delta_0 = \Delta_1 = 0$ then both $C_+$ and $C_-$ are zero. In this case, all three roots are given by
\begin{equation}
    \label{eq:Cubic 3 solutions}
    w_0 = \frac{-b}{3a},
\end{equation}
which is always real.

\subsection{Summary}

The real roots of the cubic Eq.~\ref{eq:General cubic} are tabulated in Table~\ref{tab:cubic_real_roots}. For cases where we have multiple real roots, we must determine which of these gives the global maximum of the likelihood.

\begin{table}
    \centering
    \begin{tabular}{c|c|c}
    & $\Delta_1^2 \leq 4 \Delta_0^3$ & $\Delta_1^2 > 4 \Delta_0^3$  \\
    \hline
    $\Delta_1 + \sqrt{\Delta_1^2 - 4 \Delta_0^3} \neq 0$ & $u_0$, $u_1$, $u_2$ & $u_0$\\
    $\Delta_1 - \sqrt{\Delta_1^2 - 4 \Delta_0^3} \neq 0$ & $v_0$, $v_1$, $v_2$ & $v_0$ \\
    $\Delta_0 = \Delta_1 = 0$ & $w_0$ & $w_0$
    \end{tabular}
    \caption{The real solutions of Eq.~\ref{eq:General cubic}. $u_k$ is defined in Eq.~\ref{eq:Cubic 1 solutions}, $v_k$ in Eq.~\ref{eq:Cubic 2 solutions} and $w_k$ in Eq.~\ref{eq:Cubic 3 solutions}.}
    \label{tab:cubic_real_roots}
\end{table}

Upon closer inspection, it is possible to eliminate some of these choices. If we say that our default set of solutions are $\{u_0, u_1, u_2\}$, then we will use these unless $C_+ = 0$. In that case, if $C_- \neq 0$, then we use the set of solutions $\{v_0, v_1, v_2\}$, and if $C_+ = C_- = 0$, then our only solution is $w_0$. But, if $C_+=0$, this means $\Delta_1 = - \sqrt{\Delta_1^2 - 4 \Delta_0^3}$. Since $\Delta_0, \Delta_1 \in \mathbb{R}$, this means $\Delta_1^2 \geq 4 \Delta_0^3$. Looking at Table~\ref{tab:cubic_real_roots}, this means that we have only one real root, except perhaps in the case where $\Delta_1^2 = 4 \Delta_0^3$. But this case then requires $\Delta_1 = 0$ (so that $C_+ = 0$), and hence $\Delta_0=0$. But this gives $C_- = 0$, in which case we do not use this set of solutions, and instead use $w_0$.

The procedure for computing the real roots of Eq.~\ref{eq:General cubic} is given in Algorithm~\ref{alg:cubic real roots}. There is now only one case where we have three real roots to consider, which is $C_+ \neq 0$ and $\Delta_1^2 \leq 4 \Delta_0^3$.

\begin{algorithm}
\caption{Algorithm to compute the real roots of the cubic $au^3 + bu^2 + cu + d = 0$. Here $\atantwo(y,x)$ is the usual function to give the argument of the complex number $x+\iu y$. See Section \ref{sec:cubic properties} for a derivation of this algorithm.}\label{alg:cubic real roots}
\begin{algorithmic}
\State $\Delta_0  \gets b^2 - 3ac$
\State $\Delta_1  \gets 2 b^3 - 9abc + 27 a^2 d$ 
\State $C_+ \gets \left( \frac{\Delta_1 + \sqrt{\Delta_1^2 - 4 \Delta_0^3}}{2} \right)^{1/3}$
\If{$C_+ \neq 0$} 
    \If{$\Delta_1^2 > 4 \Delta_0^3$}
        \State $u \gets - \frac{1}{3a} \left( b + C_+ + \frac{\Delta_0}{C_+} \right)$
    \Else
        \State $\theta_+ \gets  \frac{1}{3} \atantwo \left( \sqrt{4 \Delta_0^3 - \Delta_1^2}, \Delta_1 \right)$
        \State $u_0 \gets - \frac{1}{3a} \left( b + 2\Delta_0^{1/2} \cos \theta_+\right) $
        \State $u_1 \gets - \frac{1}{3a} \left( b + 2\Delta_0^{1/2} \cos  \left( \theta_+ + \frac{2 \pi}{3} \right) \right) $
        \State $u_2 \gets - \frac{1}{3a} \left( b + 2\Delta_0^{1/2} \cos \left( \theta_+ - \frac{2 \pi}{3} \right)  \right) $
        \State $u \gets [u_0, u_1, u_2]$
    \EndIf
\Else
    \State $C_- \gets \left( \frac{\Delta_1 - \sqrt{\Delta_1^2 - 4 \Delta_0^3}}{2} \right)^{1/3}$
    \If{$ C_- \neq 0$} 
        \State $u \gets - \frac{1}{3a} \left( b + C_- + \frac{\Delta_0}{C_-} \right)$
    \Else
        \State $u \gets - \frac{b}{3a}$ 
    \EndIf
\EndIf 
\end{algorithmic}
\end{algorithm}

\label{lastpage}
\end{document}